\documentclass[10pt, aps,longbibliography,superscriptaddress, twocolumn, pra, english]{revtex4-2}

\usepackage{bm}
\usepackage{bbold}
\usepackage{natbib}

\usepackage[breaklinks=true, colorlinks =true]{hyperref}
\usepackage{amsfonts,amsxtra}
\usepackage{graphicx}
\usepackage{epsfig}
\usepackage{blindtext}
\usepackage{grffile}
\usepackage{amsfonts}
\usepackage{placeins}
\usepackage[linesnumbered,ruled,vlined]{algorithm2e}
\SetKwFunction{ILP}{ILP}
\SetKwProg{Fn}{Function}{}{end}
\usepackage{LatexCommands}
\usepackage[most]{tcolorbox}
\usepackage{floatrow}
\usepackage{verbatim}
\usepackage{color}
\usepackage{diagbox}
\usepackage{ulem}
\usepackage{cancel}
\usepackage[ruled,vlined]{algorithm2e}
\usepackage[noend]{algpseudocode}

\providecommand{\abs}[1]{\lvert#1\rvert}

\newcommand{\intersection}{\cap}

\newcommand{\CP}[1]{\textcolor{red}{[#1]}}
\usepackage[english]{babel}
\providecommand{\selectlanguage}[1]{}

\allowdisplaybreaks

\begin{document}


\title{Leveraging Analog Neutral-Atoms Quantum Computers for Diversified Pricing in Hybrid Column Generation Frameworks}

\author{Cédrick Perron}
\affiliation{Institut quantique, Sherbrooke, Québec, J1K 2R1, Canada}
\affiliation{Département de génie électrique et de génie informatique, Université de Sherbrooke, Sherbrooke, Québec, J1K 2R1, Canada}
\author{Yves Bérubé-Lauzière}
\affiliation{Institut quantique, Sherbrooke, Québec, J1K 2R1, Canada}
\affiliation{Département de génie électrique et de génie informatique, Université de Sherbrooke, Sherbrooke, Québec, J1K 2R1, Canada}
\author{Victor Drouin-Touchette}
\affiliation{Institut quantique, Sherbrooke, Québec, J1K 2R1, Canada}
\affiliation{Département de génie électrique et de génie informatique, Université de Sherbrooke, Sherbrooke, Québec, J1K 2R1, Canada}
\affiliation{Département de physique, Université de Sherbrooke, Sherbrooke, Québec, J1K 2R1, Canada}
\date{\today}

\begin{abstract}
In this work, we develop new pulse designs and embedding strategies to improve the analog quantum subroutines of hybrid column generation (CG) algorithms based on neutral-atoms quantum computers (NAQCs). These strategies are designed to improve the quality and diversity of the samples generated. We apply these to an important combinatorial optimization (CO) problem in logistics, namely the fleet assignment.  Depending on the instance tested, our quantum protocol has a performance that is either comparable or worse than the best classical method tested, both in terms of the number of iterations and final objective value. We identify the cause of these suboptimal solutions as a result of our quantum protocol often generating high-quality but degenerate samples. We address this limitation by introducing a greedy post-processing technique, Make\_Diff, which applies bit-wise modifications to degenerate samples in order to return a non-degenerate set. With this modification, our quantum protocol becomes competitive with an exact solver for the subproblem, all the while being resilient to state preparation and measurements (SPAM) errors. We also compare our CG scheme with a Gurobi solver and find that it performs better on over 50\% of our synthetic instances and that, despite Gurobi having a more extensive runtime. These improvements and benchmarks herald the potential of deploying hybrid CG schemes on NISQ devices for industrially relevant CO problems.

\end{abstract}
\maketitle

\section{Introduction}
Combinatorial optimization (CO) problems involve selecting the optimal solution from a finite, yet exponentially large set of possible solutions subject to constraints. They are ubiquitous in domains such as logistics, scheduling, and resource allocation, where even marginal improvements in optimization can yield substantial operational and economic benefits. Quantum algorithms have thus attracted attention for their potential to speedup certain classes of these problems \cite{leng_subexponential_2025, huang_quantum_2022, montanaro_quantum_2020, wocjan_speedup_2008, moylett2017quantum, ambainis2019quantum, somma_quantum_2012}, which would amplify the operational and economic gains. Nonetheless, current industrial implementations are rare.

While quantum computing offers alternative algorithmic strategies based on entanglement and superposition, it operates in a large Hilbert space of size $O(2^N)$, with $N$ the number of qubits in a quantum system. Quantum algorithms must therefore manipulate quantum information in this space, crafting interference patterns so as to concentrate probability on the answer. An example of this recipe for unstructured search is Grover's algorithm \cite{montanaro_quantum_2020}, which offers a quadratic speedup over a brute force search \cite{creemers_speeding_2025}, though additionnal limitations may be present \cite{creemers_perez_armas_limitations_2025}. 
Such digital algorithms for CO mostly remain out of reach in the noisy intermediate-scale quantum (NISQ) regime due to the prohibitively large number of coherent operations they require. 

Quantum annealing approaches, inspired by the adiabatic theorem \cite{Albash_2018, rajak2023quantum, hauke2020perspectives}, have been explored extensively as more NISQ-friendly protocols \cite{yarkoni_quantum_2021, mohammed_review_2025, kim_quantum_2025, perez_armas_creemers_deleplanque_solving_RCPSP_QA_2024}. They typically start by representing an optimization problem as a quadratic unconstrained binary optimization (QUBO), where constraints become slack variables and Lagrange multipliers \cite{Lucas_2014, glover2018tutorial}. Through the annealing, one carries a known product state under a time-varying Hamiltonian such that the final Hamiltonian encodes the cost function (i.e. the QUBO). In the adiabatic limit, the final state will be the solution to the QUBO, although that limit is generally unachievable for NISQ devices with a finite coherence time. Instead, annealing is done in a finite time, and one typically encounters a very small energy gap which leads to the proliferation of defects \cite{king2022coherent, soriani2022three}. Even in a fully coherent system, these defects may lead to observed states that are not even feasible solutions \cite{cazals2025quantum}. The quantum approximate optimization algorithm (QAOA) \cite{farhi2014quantumapproximateoptimizationalgorithm, Wurtz:2021uec}, a heuristic method designed to approximate optimal annealing paths on digital quantum computers, suffers from similar problems, although methods such as subspace embedding \cite{QAOA+1, QAOA+2} can remedy them, at the cost of deeper circuits.

Classical algorithms, such as branch-and-price-and-cut~\cite{desrosiers2024branch}, harness the structure of the constrained optimization problem for increased performance.
They use it to refine heuristics, find cuts, generate new variables, set new bounds. 
On the other hand, quantum approaches such as those presented above rely mostly on unstructured search.
In this dichotomy lies the need for quantum algorithms that take advantage of efficient mathematical formulations, while tackling the biggest bottlenecks with new quantum protocols. Quantum column generation (QCG) schemes \cite{da_silva_coelho_quantum_2023, chatterjee_hybrid_2024, perez_armas_hybrid_2025, hirama_efficient_2023, huang_solving_2025} fill this need.
They leverage the classical column generation (CG) framework, where the problem is split into an easily solvable master problem and hard subproblems with a common structure. The subproblems focus on identifying new columns (or variables) that may improve the master problem's optimum. These subproblems are themselves hard CO problems that are solved iteratively many times by heuristics, representing a major  bottleneck for performance. Initial results suggest that quantum approaches to the subproblems can lead to better overall results than other heuristics, such as genetic algorithms \cite{chatterjee_hybrid_2024}, simulated annealing \cite{hirama_efficient_2023, da_silva_coelho_quantum_2023}, and even exact methods \cite{hirama_efficient_2023, da_silva_coelho_quantum_2023}. 

In some particular cases, such as the graph coloring problem, the subproblems' structure as maximum weighted independent sets (MWIS) instances makes them particularly amenable to neutral-atoms quantum computers (NAQC) \cite{da_silva_coelho_quantum_2023}. This is because the analog and continuous control of these interacting many-body systems can reach a dynamical regime dominated by the Rydberg blockade, which forbids atoms closer than a certain distance from being both in their respective excited states. This is exactly the independence constraint on edges in a MWIS instance: $n_i + n_j \leq 1$ ($n_i$ is a projector on the Rydberg excited state). Another practically important CO problem, the fleet assignment problem, in which one seeks to assign tours to vehicles in a way that satisfies operational constraints and minimizes the total cost of operations, can also be tackled by identifying the same structure in the subproblems~\cite{chatterjee_hybrid_2024}, making it also amenable to analog protocols on NAQC. 

One key factor in accelerating CG is the ability of the heuristic to quickly produce, at each iteration, a set of many, good, and diverse columns \cite{aor5_2art03, desauliers2005column}. Therefore, for quantum heuristics to offer meaningful advantage over classical ones, they must demonstrate performance in generating a set of columns under these criteria. There are encouraging signs in this direction with prior studies suggesting that quantum annealing may be more effective than its classical counterpart at sampling diverse solutions \cite{zucca2021diversity, dziubyna2024limitations}. Furthermore, quantum heuristics on NAQCs for the MWIS problem have shown promising performance in the solution quality \cite{ebadi_quantum_2021, cazals2025identifying}.
This raises two questions. Firstly, can quantum heuristics be designed that specifically return a set of many, good, and diverse solutions? Secondly, when implemented in a QCG framework, do these quantum heuristics perform better, identically, or worse than competing classical methods? This contribution seeks to address these two questions.



\begin{figure*}[!t]
\centerline{
\includegraphics[width=0.9\linewidth]{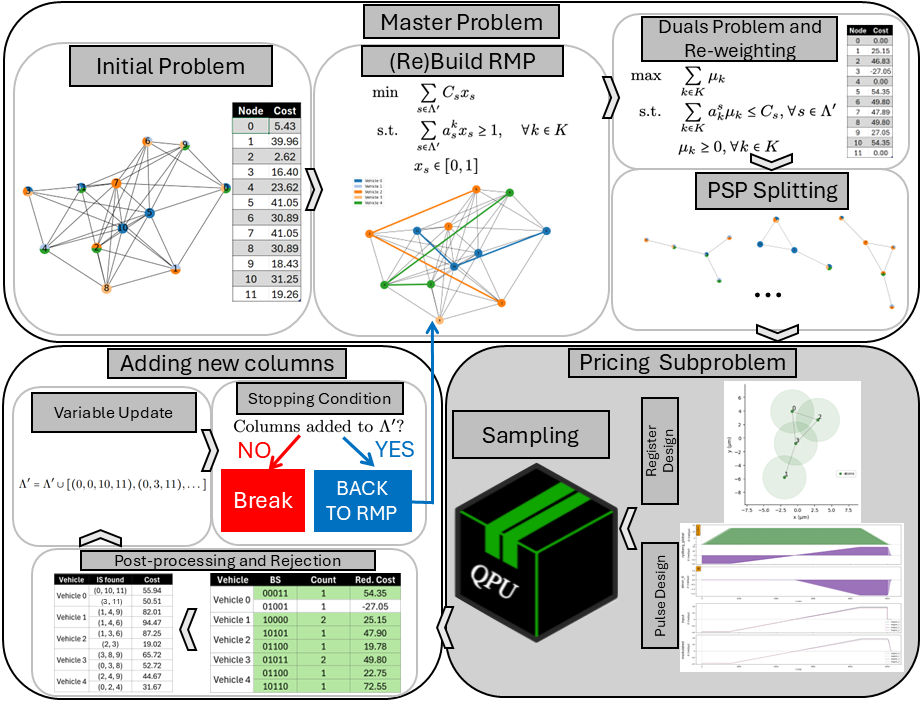}}
 \caption{Workflow of the tested hybrid classical-quantum column generation approach for the fleet assignment problem, which uses a neutral-atoms quantum computer (NAQC) as a sampler of many, good, and diverse potential variables to add to the classical workflow. In the top left, a fleet assignment problem is represented as a conflict graph; nodes represent available tours, colors compatible vehicle classes, and the cost table represents the tour costs (vehicle costs not shown). After building the RMP with an initial set of columns $\Lambda'$, one obtains its optimal dual solution, which serves to construct a series of maximum weighted independent set problems (called pricing subproblems, PSPs) that are each sent to the NAQC. There, we use our register and pulse design strategies to tailor the dynamics in order to measure a set of good and diverse independent sets for the PSPs. These are post-processed, and then the ones passing a threshold are accepted in the column set $\Lambda'$. The process terminates when no improving column is found. An example of the primal solution after a single CG iteration is shown under '(Re)Build RMP'.}
 \label{fig:workflow}
\end{figure*}

To tackle these questions, we sought a hard CO problem on which to perform benchmarks. We chose the fleet assignment problem for its added benefit of being highly relevant to the logistics industry. We construct a QCG framework inspired from Ref.~\cite{da_silva_coelho_quantum_2023, chatterjee_hybrid_2024} for the fleet assignment problem. The workflow is shown in Fig.~\ref{fig:workflow}, and described in detail in Sec.~\ref{sec:colgen}. 
Expanding on the analog compilation introduced in \cite{da_silva_coelho_quantum_2023}, we introduce a new classical algorithm for embedding a given MWIS graph instance into an spatial arrangement of atoms (a register) and a Rydberg blockade radius. Our algorithm based on simulated annealing leads to embedded graphs that have fewer extra edges and are thus more easily realized on the hardware. We also propose a new pulse design which favors quantum fluctuations towards the end of the annealing and thus increases diversity in the sampled bitstrings. We validate these insights on random graph instances. We show later on our benchmarks on synthetic instances that these changes to the analog compilation, which aim to increase the overall quality and diversity of the samples obtained, result in higher performances. 

We then compare our analog quantum protocol with a variety of exact and heuristic classical methods that can also return many solutions to the subproblem. This helps separate the effect of column intensification, where more than one column is added at each iteration, from the quality and diversity objectives. We show that these classical methods, in particular simulated annealing, generally achieve similar/stronger performance, both in the number of iterations and the final objective value. This is because these classical methods almost never generate degenerate columns, thus we introduce a new and fast postprocessing step called Make\_Diff which guarantees unique potential columns. We show that this routine particularly helps our quantum protocol, whose performance becomes competitive with an exact solver for the subproblem.


The paper is organized as follows. After introducing the fleet assignment problem in Sec.~\ref{sec:fleetass}, we cover the column generation (CG) decomposition for this problem in Sec.~\ref{sec:colgen}, as well as several classical approaches to the subproblems. In Sec.~\ref{sec:analog}, we cover the fundamentals of neutral-atoms quantum computers, introducing our novel register embedding algorithm (Sec.~\ref{sec:embed}) and pulse design (Sec.~\ref{sec:pulse}). Our results are presented in Sec.~\ref{sec:results}, with a particular emphasis on the performance of the different protocols for the subproblems with respect to quality and diversity of the columns returned (Sec.~\ref{sec:quality}). Finally, we offer concluding remarks and final insights into the conditions under which this quantum workflow may yield a practical advantage.

\section{Fleet Assignment} \label{sec:fleetass}
The fleet assignment problem is a complex combinatorial optimization problem that involves assigning a set of vehicles to a set of transportation tasks (called tours), in a way that minimizes the total operational cost. Each tour corresponds to a delivery, route, or service task defined over a time window. Furthermore, the fleet of vehicles is generally heterogeneous, consisting of multiple vehicle classes, each with its own cost structure, operational limitations, and availability bounds (often specified in terms of minimum and maximum allowable vehicles per class). Thus, not all tours might be compatible with all vehicle classes, which gives rise to restrictions due to capacity, regulation, or operational compatibility constraints. In industrial settings, the fleet assignment problem is typically embedded within a larger integrated problem, such as airline scheduling, where multiple subproblems such as crew scheduling, routing, and maintenance planning must be solved in combination \cite{airline-fleet-assignment, airline-fleet-assignment-2}. In this paper, we focus on only the fleet assignment component of such larger settings.

Mathematically, we define the following objects. Let $K = \{k_1, \ldots, k_{|K|}\}$ be a set of $|K|$ tours. Each tour $k_i$ is associated with a tour cost $C^{(T)}_{k_i}$, a time window $T_{k_i} = [t^{k_i}_{\text{start}}, t^{k_i}_{\text{end}}]$ during which the tour must be traversed, and a set of compatible vehicle classes $V_{k_i} \subseteq V$, where $V = \{v_1, \ldots, v_{|V|}\}$ is the set of all $|V|$ vehicle classes. Each vehicle class $v_j$ has an associated operational cost $C^{(v)}_{v_j}$ and availability bounds $N_{v_j} \in [N_{v_j}^{\min}, N_{v_j}^{\max}]$. 


To better grasp the relationship between these objects, we introduce a conflict graph $G = (K, E)$, where each node $k \in K$ corresponds to a tour as previously defined. An edge $e_{kk'} \in E$ is added between two nodes $k$ and $k'$ if 1) the corresponding tours cannot be assigned to the same vehicle due to overlapping time windows ($(t^{k}_{\text{start}} < t^{k'}_{\text{end}}) \wedge (t^{k}_{\text{end}} < t^{k'}_{\text{start}})$) or 2) incompatible vehicle classes ($V_k \cap V_{k'} = \emptyset$). Other constraints could lead to more edges being added to the conflict graph. Therefore, any subset of the nodes $\tilde{K} \subset K$ such that not two nodes of $\tilde{K}$ share an edge is a feasible vehicle-to-tour assignment. This is exactly the definition of an independent set on the conflict graph $G$. We represent an independent set as the binary vector $\Vec{a} = \{a_1, ..., a_{|K|}\}$, where $a_k = 1$ denotes the presence of the route $k$ in the independent set and $0$ denotes its absence. We denote $\Omega(G)$ as the set of all independent sets in $G$. For a given independent set $\Vec{a}$, one also has to assign the compatible vehicle fleet $v$.
We denote a specific assignment with another binary vector $\Vec{b} = \{b_1, ..., b_{|V|}\}$ where $b_v = 1$ if $v$ is the vehicle type chosen and $0$ otherwise, such that $\sum_{v=1}^{|V|} b_j = 1$, i.e. only one vehicle class has been assigned to this set of routes. Together, the set of all vehicle-to-tour assignments, denoted by $\Lambda$, can formally be expressed as:

\begin{equation}
\begin{split}
    \Lambda = \{ (\Vec{a}, \Vec{b}) \; \lvert &\; (\Vec{a} \in  \Omega(G)) \; \land \; \\
    &b_v = 1 \; \text{if} \; v \in V_k \; \forall k \; \text{such that} \; a_k = 1 \}
    \end{split}
\end{equation}

A given assignment $\lambda_s = (\Vec{b}_s, \Vec{a}_s) \in \Lambda$ can thus be represented using a binary vector of size $|K||V|$. We then introduce $x_s$ as a binary variable indicating whether the vehicle-to-tour assignment $\lambda_s$ is selected (1) or not (0), and $C_s$ as the total cost of the assignment, which can be expressed as




\begin{equation}
C_s = \sum_{v=1}^{|V|} b_{v,s} C^{(v)}_v + \sum_{k=1}^{|K|} a_{k,s} C^{(T)}_k \;,
\end{equation}

where $b_{v,s}$ is the $v$-th component of the vector $\Vec{b}_s$ (idem for $a_{k,s}$ and $\Vec{a}_s$). The fleet assignment problem, corresponding to the task of finding the best assignments $\lambda_s$ to select so as to minimize operational cost while satisfying the constraints, can then be formulated as the following integer linear program (ILP):
\begin{equation}
\begin{split}
    \min \quad & \sum_{\lambda_s \in \Lambda} C_s x_s \label{eq:IMP} \\
    \text{s.t.} \quad & \sum_{\lambda_s \in \Lambda} a_{k,s} x_s \geq 1, \quad \forall k \in K  \\
                      & N_v^{\min} \leq \sum_{s \in \Lambda} b_{v,s} x_s \leq N_v^{\max}, \quad \forall v \in V  \\
                      & x_s \in \{ 0, 1 \} ,
\end{split}
\end{equation}


The first constraint enforces the visitation of each tour $k \in K$. This constraint allows the overlap of tours (meaning more than one vehicle can be assigned to the same tour). It can be easily turned into an equality which would force each tour to be undertaken by only one vehicle. We have reintroduced in the second constraint the $N_{v}^{\min}$ and $N_v^{\max}$ constraints, which are respectively the minimum and maximum number of vehicles that is allowed in class $v$. By choosing $N_v^{\min} = 0$ and $N_v^{\max} = 1$, we can transform the meaning of $v$ from a vehicle class to an individual vehicle. The last line is the binary constraint on the variables. 


The challenge of solving the fleet assignment problem is evident from the problem stated above. The size of the set $\Lambda$ increases exponentially in both the number of vehicle classes $|V|$ and of tours to service $|K|$. Thus, navigating such a large set of potentially feasible solutions is the bottleneck that makes brute-force approaches impractical. In practical settings, this problem is often modeled as a mixed-integer linear program (MILP) (see Ref.~\cite{chatterjee_hybrid_2024} for an example formulation), where the number of assigned vehicles per class is an unbounded positive natural number. State-of-the-art approaches to MILPs include branch-and-bound, column and row generation and bender's decompositions~\cite{Fleet-assignment-BB, Fleet-assignment-state-of-the-art}. These techniques are sometimes further improved by greedy heuristics. The software Gurobi~\cite{gurobi} implements most of these approaches and is among the best black-box solvers of MILPs, thus we use it to benchmark the hardness of our tested instances. In the rest of the paper, we explore the column generation framework to solve the fleet conversion problem, and describe how this leads to a series of subproblems to solve that are especially compatible with analog neutral-atoms quantum computers (NAQCs).

\section{Column Generation} \label{sec:colgen}


Column generation (CG) is a powerful optimization technique used to efficiently solve large-scale integer linear problems (ILP). It is particularly relevant when the problem involves an exponential explosion in the number of feasible solutions, such as in the ILP presented in Eq.~\eqref{eq:IMP}. Explicitly considering the entire set $\Lambda$ becomes computationally intractable, but CG offers a solution by iteratively generating the most promising variables (columns) that will improve the solution, thus constructing a restricted set $\Lambda'$ such that $|\Lambda'| \ll |\Lambda|$. Then, the exact solution of the restricted ILP over $\Lambda'$ is possible due to its now very reasonable size. This makes CG a highly effective tool for tackling complex, large-scale problems in fields such as logistics, telecommunications, and transportation \cite{desauliers2005column, Des-VRP-Col-Gen, Des-Scheduling-Col-Gen}. 

At the core of CG is the decomposition of the main optimization problem into two subproblems: a restricted master P
problem (RMP) and the pricing sub-problem (PSP). The RMP is a relaxed version of the original ILP, which uses a limited subset of variables (columns) and where the binary constraint of the variables is dropped (i.e. $0 \leq x_s \leq 1$). Linear programming (LP) solvers, such as CPLEX and GLPK, can then be used to return a solution in quasi-polynomial time. One then uses this solution, and in particular the value of the dual variables at optimality \cite{Duality-Theory}, to construct a pricing sub-problem (PSP). Near-optimal solutions of the PSP, should they pass a certain threshold, are guaranteed to correspond to useful columns (e.g. new tour-vehicle assignments) to add to $\Lambda'$. This loop is repeated until no further columns can be found to improve the RMP (e.g. decrease the cost of the fleet assignment), signaling that the optimal solution has been reached. The PSP is often the bottleneck of the CG approach due to the computational complexity of solving the underlying optimization problem, which can itself be NP-Hard and, in the worst case, APX-Hard, meaning that even good approximations are hard to obtain in polynomial time. Additionally, the quality and diversity of the columns generated in the PSP play a crucial role in ensuring fast convergence and optimal solutions. Poorly chosen columns can result in slow convergence, suboptimal solutions, and inefficiency in the optimization process.

\subsection{Reduced Master Problem}

The RMP for our fleet assignment problem is given by first considering a limited subset of columns $\Lambda' \subset \Lambda$ in Eq.~\eqref{eq:IMP}. This set is kept small such that the solution to this new ILP can be done quickly (this is usually on the order of $|\Lambda'| \approx O(10^2)$ variables). Usually, this set will be formed, at iteration $0$, of a minimal subset of singletons solutions (e.g. vehicle-tour assignment with a single tour per vehicle) that ensures that the first constraint of Eq.~\eqref{eq:IMP} is satisfied \footnote{Another approach taken by Ref.~\cite{chatterjee_hybrid_2024} is to add a relaxation variable $r_k$ which becomes zero when the tour is not executed by any column. When this happens, a penalty cost $R \cdot r_k$ where $R \gg \max{(C_k + C_v)} $ is added.}. We then relax the constraint on the binary variables $x_s$, allowing them to become real numbers such that $0 \leq x_s \leq 1$. We obtain:
\begin{equation}
\begin{split}
    \min \quad & \sum_{\lambda_s \in \Lambda'} C_s x_s \label{eq:primal-cost} \\
    \text{s.t.} \quad & \sum_{\lambda_s \in \Lambda'} a_{k,s} x_s \geq 1, \quad \forall k \in K \quad [\mu_k]  \\
                      & \sum_{\lambda_s \in \Lambda'} b_{v,s} x_s \geq N_v^{\min} \quad \forall v \in V \quad [\mu_v^{\min}] \\
                      & \sum_{\lambda_s \in \Lambda'} b_{v,s} x_s \leq N_v^{\max} \quad \forall v \in V \quad [\mu_v^{\max}]  \\
                      & 0 \leq x_s \leq 1 \quad \forall s \in \Lambda'
\end{split}
\end{equation}

From duality theory of LPs \cite{Duality-Theory}, we know that each constraint has an associated dual variable, which we have added in bracket next to their respective constraint. The dual problem is thus a maximization problem on the dual variables, and reads as:
\begin{equation}
\begin{split}
    \max \quad & \sum_{k \in K} \mu_k + \sum_{v \in V} \left( N_v^{\min} \mu_v^{\min} + N_v^{\max} \mu_v^{\max} \right) \label{eq:dual-cost} \\
    \text{s.t.} \quad & \sum_{k \in K} a_{k,s} \mu_k                              \leq C_s,  \forall s \in \Lambda' \quad [x_s]  \\
                      & \sum_{v \in V} b_{v,s} \mu_v^{\min}                       \leq 0,    \forall s \in \Lambda'  \\
                      & \sum_{v \in V} b_{v,s} \mu_v^{\max}                       \geq 0,   \forall s \in \Lambda'  \\
                      & \mu_k                                                \geq 0,    \forall k \in K 
\end{split}
\end{equation}

LP solvers solve both the primal (Eq.~\eqref{eq:primal-cost}) and the dual (Eq.~\eqref{eq:dual-cost}) simultaneously. When their respective optimal values meet, the solution to the LP is optimal. This solution is not necessarily integer in the values $x_s$, although it will be if the set $\Lambda'$ contains the columns forming the optimal integer solution.

\subsection{Pricing Sub-Problem}

Once a solution to the LP is obtained with the limited set $\Lambda'$, one can formulate the pricing sub-problem (PSP) whose solution will lead to new columns to augment $\Lambda'$. The formulation of the PSP is derived from the constraints of the dual formulation of Eq.~\eqref{eq:dual-cost}, which can be combined to give:
\begin{equation}
    \sum_{k \in K} a_{k,s} \mu_k + \sum_{v \in V} b_{v,s} ( \mu_v^{\max} - \mu_v^{\min}) \leq C_s \text{, } \forall s \in \Lambda' \;.\label{PSP-bigconstraint}
\end{equation}
where that $\mu_k$, $\mu_v^{\min}$ and $\mu_v^{\max}$ are the dual solutions to the LP. The goal of CG is to find new columns $\lambda_{s'} = (\Vec{b}_{s'}, \vec{a}_{s'})$ for which the above inequality does not hold, indicating that the current solution is not yet optimal. The addition of these new columns in the primal problem amounts to the addition of new constraints in the dual problem. Therefore, from the perspective of the dual problem, the PSP searches for a column $\lambda_{s'}$ whose coefficients $a_{k,s'}$ and $b_{v,s'}$ maximize the violation of the above inequality (i.e. maximize the left-hand side) under the current dual solution ($\mu_k$, $\mu_v^{\min}$, $\mu_v^{\max}$). These coefficients also need to respect the constraints of the primal problem (Eq. \eqref{eq:primal-cost}). Thus, the PSP can be written as a maximization problem:
\begin{equation}
\begin{split}
    \max \sigma = & \max \sum_{k \in K} n_k \left( \mu_k - c_k \right) \\
    &+ \sum_{v \in V} m_v \left( \mu_v^{\min} + \mu_v^{\max} - c_v \right)   \\
    \text{s.t.} \quad 
    & n_k + n_{k'} \leq 1, \quad \forall \; e_{kk'} \in E \\
    & \sum_v m_v \leq 1 \; \\
    & n_k, m_v \in \{0,1\} \; \forall k \in K, v \in V\label{PSP-cost}
    \end{split}
\end{equation} 
where we replace coefficients $a_{k,s'}$ and $b_{v,s'}$ by $n_k$ and $m_v$ respectively. Note that we drop the subscript $s'$ since the PSP focuses on finding a single candidate column. Any solution tuple $(\Vec{m}^{\ast}, \Vec{n}^{\ast})$ with cost $\sigma^{\ast}$ where $\sigma^{\ast} > 0$ then represents a new violation of the inequality of Eq.~\eqref{PSP-bigconstraint}. It is possible, and in fact likely, that multiple such columns exist. It is then added to the set of columns $\Lambda' \rightarrow \{\lambda_{\ast}\} \bigcup \Lambda'$.


The constraint in Eq.~\eqref{PSP-cost} corresponds to the independent set (IS) constraint on the conflict graph $G = (K, E)$. Thus, this PSP is entirely specified by the initial graph $G$ of the problem and the dual variables of the previous iteration. This PSP is a modified maximum weighted independent set (MWIS) problem, with the added complexity of finding the appropriate vehicle class $v$ (which $m_v$ is equal to one) for a given IS (a vector $\vec{n}$). This can be tackled by decomposing this PSP in $|V|$ problems, called PSP${}_v$ which each seek to find new columns compatible with each vehicle class $v \in V$. 


This can be achieved by defining a reduced graph $G_v = (K_v, E_v)$, a subgraph of the total graph $G = (K, E)$ containing only the tours that are feasible for a vehicle type $v$. We have $K_v = \{ k \in K \lvert v \in V_k \}$, and $E_v = \{ e_{kk'} \in E \lvert k, k' \in K_v \}$. Then, the PSP${}_v$ is expressed as
\begin{equation}
\begin{split}
    \max \sigma_v = & \max \sum_{k \in K_v} n_k \left( \mu_k - c_k \right) \\
    \text{s.t.} \; & \; n_k + n_{k'} \leq 1, \quad \forall \; e_{kk'} \in E_v \\
    & n_k \in \{0,1\} \; \forall k \in K_v\label{subPSP-cost}
    \end{split}
\end{equation} 

The acceptance condition for a new column for the vehicle class $v$ becomes:
\begin{equation}
    \sigma_v > - ( \mu_v^{\min} + \mu_v^{\max} - c_v )
\end{equation}
If no such solution exists for any vehicle class, then the RMP cannot be improved and the optimal solution of the RMP is reached by solving the binary RMP with the obtained set of columns $\Lambda'$, i.e. replacing the constraint $0 \leq x_s \leq 1$ by $x_s \in \{0,1\}$. Alternatively, if columns are added to the RMP at each iteration but the relaxed optimum does not change, we stop the process and solve the binary RMP - this can return degraded (non-optimal) fleet assignment solutions.


As in Ref.~\cite{da_silva_coelho_quantum_2023}, which present a quantum approach to the column generation formulation of the graph coloring problem, we recognize that the PSP${}_v$ from Eq.~\eqref{subPSP-cost} are MWIS problems, with node weights $\omega_k = \mu_k - c_k$, $k\in K_v$, on the graphs $G_v = (K_v, E_v)$. Thus it can be readily implemented on current NAQCs by harnessing the Rydberg blockade phenomena to embed the strong independence constraints between adjacent nodes. We detail our improvements on the analog algorithm for sampling from the nearly-optimal solutions to the PSP in Sec.~\ref{sec:analog}.

\subsection{Classical approaches for the PSP} \label{sec:classical}

For general conflict graphs $G$ and their associated subgraphs $G_v$, one has that the PSPs are hard to approximate, since the MWIS not only NP-Complete, but also APX-Hard \cite{cazals2025quantum}. This means that, for graphs with no particular structure, there is no polynomial classical algorithm to approximately solve this problem. This constitutes the bottleneck of the column generation algorithm. In this section, we describe three classical methods we used to solve the PSP in our benchmark instances.

Firstly, we implement an integer linear problem solver (GLPK) to exactly solve the PSP. In the method we call 1-ILP, we exactly solve the PSP${}_v$ for each vehicle class $v$. In ILP+DIV, we developed an algorithm that returns the $M$ best solutions by iteratively cutting the previously obtained optimum from the set of allowed solutions (i.e. adding a constraint to the MWIS). This algorithm is also applied to each vehicle type $v$. Details for ILP+DIV are provided in Appendix \ref{sec:Appendix-ILP+DIV}.

We also implement a greedy solver, where the probability of selecting a vertex in an independent set is determined by the cumulative probability of the (normalized) weights. By changing the random seed (initial node chosen in the IS), one can return $M$ greedy solutions.

Finally, we use the simulated annealing samplers from the \textsc{neal} Python library \cite{dwave_ocean_samplers}, to which we feed the MWIS as a QUBO. For a given graph $G = (K,E,w)$, the QUBO is given by $Q_G = - \sum_{i \in K} \bar{w}_i n_i + 1.2 \sum_{ij \in E} n_i n_j$ ($\bar{w}_i \leq 1$ are renormalized node weights). This simulated annealing sampler can be fed initial ($\beta_i$) and final ($\beta_f$) inverse temperatures. We always use $\beta_i = 0.01$. In the benchmarks bellow, we show the results for both $\beta_f = 10$, which we dub "solver" for its ability to very often return the optimal MWIS solution, and for $\beta_f = 1$, which is closer to the phase transition between the ordered and disordered phases of this QUBO. We dub this second regime the "sampler" for its propensity to return a diverse set of less-than-optimal solutions.

Note that all these solvers except 1-ILP may return many potential columns to $\Lambda'$ simultaneously, so long as their cost passes the threshold. This process is called column intensification, and is known to drastically reduce the number of column generation iterations required~\cite{desauliers2005column}. In the benchmarked classical and quantum methods (see Sec.~\ref{sec:analog}) presented in this paper, $M$ queries can be made to the sampler; we use $M=5$, so that all methods are used on the same footing.

\section{Analog Quantum Algorithm for Diversified Pricing} \label{sec:analog}

Quantum protocols to solve the PSP in CG workflows have been explored as potential heuristics that could rival classical ones. The quantum approximation optimization algorithm (QAOA), a digital quantum algorithm, has been tested in Refs.~\cite{chatterjee_hybrid_2024, svensson2023hybrid}. Quantum annealing using D-Wave's system has been studied in Refs.~\cite{ossorio2022optimization, huang_solving_2025}. A quantum protocol designed for neutral-atom quantum computers was proposed in Ref.~\cite{da_silva_coelho_quantum_2023}. We consider this implementation avenue most promising for two reasons. Firstly, while analog quantum protocols on NAQCs can be deployed in a QAOA-like manner (see Refs.~\cite{leclerc2025implementing, tibaldi2025analog}), they most often involve a set time-dependent schedule for the Hamiltonian parameters, akin to quantum annealing. This means that these protocols do not have to rely on the classical search of optimal pulse parameters as in QAOA, an optimization that is known to suffer from the exponential decay of gradients in the loss function \cite{mcclean2018barren, holmes2022connecting, larocca2022diagnosing} and can require many shots/iterations. Secondly, analog quantum protocols, such as quantum annealing, can be implemented on NAQCs on timescales at or near the quantum system's coherence time. Typically, in neutral atoms, $T_{\text{coh.}} \sim 4.5 \mu s$ \cite{quera_device, PasqalOrion}), thus ensuring that the measurement outcomes are those of the prepared quantum state. As a comparison, D-Wave's standard annealing protocol occurs over timescales several orders of magnitude larger than $T_{\text{coh.}}$ \cite{mehta2025understanding}. Thus, in seeking controllable quantum protocols to solve the PSP, we turn to analog protocols on NAQCs. 





A NAQC uses lasers to manipulate the electronic states of alkali metal atoms, typically Rb${}^{87}$. These atoms serve as qubits, with different pairs of electronic states representing the computational basis. In this context, we focus on a system where the qubit states are the ground state $\ket{g} = \ket{0} $ and a Rydberg-excited state $ \ket{r} = \ket{1}$. A spatial light modulator (SLM) and a moving tweezer are used to position the atoms in a register, i.e. positions $\{\vec{r}_i\}$ in the two-dimensional plane. The register can either be a set lattice (rectangular, triangular, etc.) or arbitrary. When set in these positions, the atoms interact with each other through a time-independent pairwise van der Waals (VdW) repulsive interaction. This gives rise to 
\begin{equation}
    H_{\text{int}} = \sum_{i < j} \frac{C_6}{r_{ij}^6} \hat{n}_i \hat{n}_j \;,\label{int-hamil-ryd}
\end{equation}

where $C_6/\hbar \simeq 2\pi \times 137 \text{GHz}\cdot \mu \text{m}^6$ is a constant set by the chosen Rydberg level~\footnote{We choose $|g\rangle = |5S_{1/2}, F=2, m_F = 2\rangle$ and $|r\rangle=|60S_{1/2}, m_J = 1/2\rangle$ which sets the value of $C_6$.}, $\hat{n}_i = (1 + \sigma_i^z)/2 = |r\rangle_i \langle r|_i$, and $r_{ij} = |\vec{r}_i - \vec{r}_j|$ is the real-space distance between atoms. One notices that the strength of these interactions decay sharply as a function of distance $r_{ij}$ between atoms, and it only affects excited Rydberg states.

The transition $|g\rangle \leftrightarrow |r\rangle$ is driven using a two-photon mechanism that can be approximated by the following effective local Hamiltonian \cite{leclerc2024quantum} ($\hbar = 1$):

\begin{equation}
    H_{i}(t) = \frac{\Omega(t)}{2} \sigma_i^x - \left[\delta(t) - \epsilon_i \delta_{DMM}(t) \right] \hat{n}_i  \label{loc-hamil-ryd}
\end{equation}

and $\sigma_i^x = |r\rangle_i\langle g|_i + |g\rangle_i\langle r|_i$. $\Omega(t)$ represents the time-dependent Rabi frequency. The total detuning on the Rydberg levels, $\delta_{\text{tot},i}(t) =  \delta(t) - \epsilon_i \delta_{DMM}(t)$, depends on a global part ($\delta(t)$) and can be changed depending on the atom $i$ through the detuning map modulator (DMM). The DMM leads to a modulation $-\epsilon_i \delta_{DMM}(t)$ of the detuning that qubit $i$ experiences. By assigning a set of weights $\epsilon_i \in [0, 1]$ to each atom, one has an effective local control of the qubits' energy levels without requiring individual addressing, which remains a severe challenges for neutral-atom hardware.

Once $N$ atoms have been set in place, cooled to their ground states and initialized to the product state $|\psi_0\rangle = |0\rangle^{\otimes N}$, then the new quantum state obtained after analog Hamiltonian evolution is given by

\begin{equation}
    |\phi(T)\rangle = \mathcal{T} \left[ \exp \left( -i\int_{t=0}^{T} H_{\text{Ryd}}(t) dt\right) \right] |\psi_0\rangle \label{eq:unitarysimple}
\end{equation}

where $\mathcal{T}$ is the time-ordering operator, and $H_{\text{Ryd}}(t) = \sum_{i=1}^{N} H_{i}(t) + H_{\text{int}}$ is the total Rydberg Hamiltonian given by the drive and the interaction parts. At the end of the quantum evolution, measurement is performed through fluorescence imaging, where a bright spot reveals the presence of an atom in the trap in the $|0\rangle$ ground state. Dark spots where there once was an atom are then inferred to be the measure of a qubit in the $|1\rangle$ state.


In the presence of the drive with a maximum Rabi $\Omega_{\max} = \max_t \Omega(t)$, a pair of atoms will undergo a dynamical effect called the Rydberg blockade~\cite{henriet_quantum_2020}. This effectively prohibits the quantum state from having any weight in the $|r_1, r_2\rangle = |1,1\rangle$ computational state. If two atoms are closer than the Rydberg blockade radius $R_b$, given by:
\begin{equation}
R_b(\Omega_{\max}) = \left( \frac{C_6}{\Omega_{\max}} \right)^{1/6} \;,
\end{equation}
then both qubits cannot simultaneously occupy the Rydberg excited state. This naturally leads to a unit-disk graph (UDG) representation, where each atom in the register is a vertex, and edges exist between vertices if the corresponding atoms are closer than the Rydberg blockade radius. On an edge $e_{ij}$, we have that $n_i + n_j \leq 1$, where $n_i$ is the measurement of the Rydberg occupation at site $i$. This directly enforces the independence constraint present in MWIS problems. This encoding of the constraint through a dynamical effect is an advantage compared to other approaches, such as the one presented in Ref.~\cite{chatterjee_hybrid_2024}, where constraints have to be encoded through penalty terms in a QUBO. Thus, in NAQCs, it is possible to guide the quantum evolution to sample the MWIS of the embedding graph to large probability. In the subsections Secs.~\ref{sec:embed} and \ref{sec:pulse}, we introduce an embedding algorithm to find placements of atoms on a fixed grid such that a pulse driving the atoms to interact strongly leads to the sampling of diverse yet high quality solutions to the MWIS.


\subsection{Embedding Strategy} \label{sec:embed}



In the analog quantum algorithm on NAQCs, one is faced with the NP-Hard task of embedding a random graph into a 2-dimensional UDG representation. Not all graphs have such a representation; for example, it is impossible to find a UDG representation for any $K_{1n}$ star graph with $n > 6$ vertices. This problem has been tackled with a variety of heuristics. The most notable is the use of the Fruchterman-Reingold algorithm, which treats the edges as springs and find an equilibrium position (see Ref.~\cite{coelho_efficient_2022}). Other schemes seek to either add a $O(\text{poly}(N))$ number of ancillas to construct logical gadgets, or a heuristic method (see Ref.~\cite{schuetz2025quantum} for a recent thorough overview of the approaches). Recently, it was shown~\cite{cazals2025quantum} that an unfortunate byproduct of gadgets is that an approximate solution, even near the optimal one, for the gadgetized graph, most often than not leads to a very poor solution for the original graph; gadgets do not preserve approximation ratios. Faced with this, we thus seek new algorithms to quickly return good register embeddings.

\begin{figure}
    \centering
    \includegraphics[width=\linewidth]{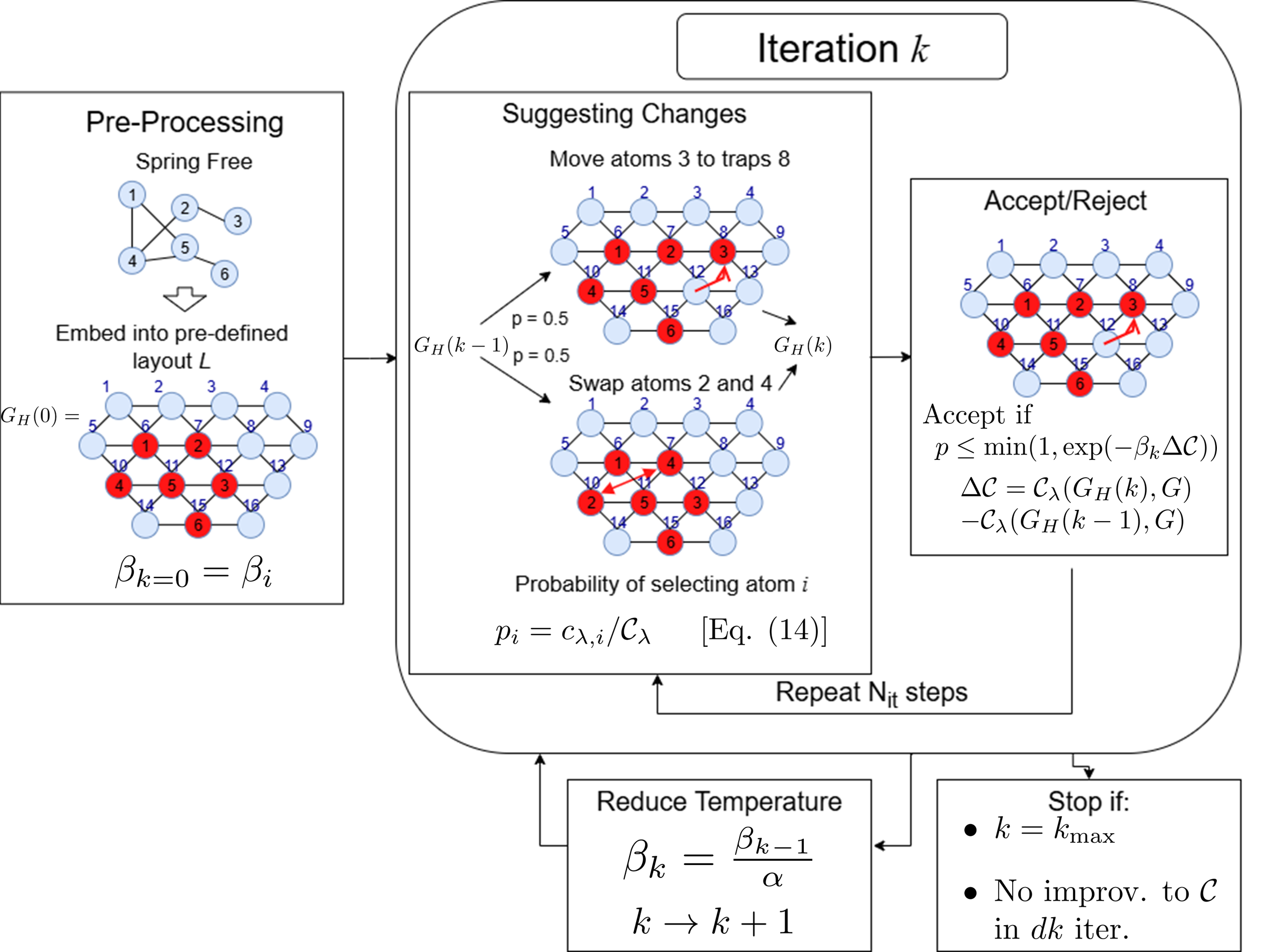}
    \caption{Schematic of \textsc{SA-Embedder}, our simulated-annealing-based protocol to embed a given graph $G$ into a user-specified layout graph $H$, resulting in an embedding graph $G_H$.}
    \label{fig:SA-workflow}
\end{figure}

We develop \textsc{SA-Embedder}, an embedding algorithm inspired by simulated annealing that moves atoms on a pre-defined layout until the best embedding is reached. A schematic view of \textsc{SA-embedder} is shown in Fig.~\ref{fig:SA-workflow}. The inputs to the algorithm are a graph $G = (V,E)$, a layout $L = \{r_i | i \in V_H\}$ corresponding to potential positions of the atoms, and a few hyperparameters controlling the cost function and the annealing path. The number of layout sites $V_H$ and the Rydberg blockade radius $R_B$ are chosen to ensure that enough sites have sufficient degree to allow for an exact UDG embedding of the graph, if one exists. We found that $|V_H| = |V|^{1.85} / \log_2(|V|)$ was sufficient. $R_B$ is then chosen so that the central layout site(s) are connected (i.e. $r_{ij} < R_B$) to at least as many layout nodes as the maximum degree of the graph. Note that in practice, a maximum $R_B$ must be set, which corresponds to a minimum $\Omega_{\max}$. Then, the combined layout $L$ and radius $R_B$ defines a UDG $H = (V_H, E_H)$. Our task is to find an assignment of nodes from $V$ into $V_H$ such that the induced subgraph $G_H$ is as close as possible to $G$; this is an embedding. To evaluate the quality of an embedding, we introduce the family of loss functions

\begin{align}
\begin{split}
    &\mathcal{C}_{\lambda} (G_H, G) = \sum_{i=1}^{V} c_{\lambda,i} \\
    &\; \; = \sum_{i = 1}^{V} \left(\sum_{i \in N_E(i)} \left[1-\mathbb{1}_{E_H}(e=(i,j)) \right]  \right.\\
    & \; \; \left.+ \lambda \sum_{i \in N_{E_H}(i)} \left[1-\mathbb{1}_{E}(e=(i,j)) \right] \right) \;,
    \label{eq:cost}
\end{split}
\end{align}

where $\mathbb{1}_A(x)$ is the indicator function that returns $1$ if $x \in A$ and $0$ otherwise, and $N_E(i)$ is the set of nodes $j$ such that $e=(i,j) \in E$. The first term in $c_{\lambda,i}$ corresponds to the number of missing edges (edges found in $G$ but not in $G_H$), while the second term corresponds to the number of extra edges (edges found in $G_H$ but not in $G$).


By tuning $\lambda$ in Eq.~\eqref{eq:cost}, one can penalize more strongly missing or extra edges. In NAQCs, limits on $R_B$ and on the minimum atomic distance make it so that there is an upper limit to the number of edges per node that can be practically realized. Thus, for general graphs with a number of edge per node commensurate with $|V|$, it is guaranteed that the embedding will have missing edges. 
In this case, one can use a fast greedy postprocessing step that explores other bitstrings that are $O(1)$ Hamming distance away from the original. For MWIS problems, this postprocessing seeks to A) fix any bitstring provided to make it independent (i.e. reintroduce missing edges) and B) greedily add nodes to the proposed independent set if they can increase the total weight (see details in Appendix~\ref{sec:Appendix-Greedy}).
On the other hand, extra edges in $G_H$ (which represent independence constraints that are not present in the original graph) leads to feasible solutions being cut from our (classical or quantum) sampling procedure.
We therefore use $\lambda = 2$ in \textsc{SA-embedder} to promote missing edges over extra edges in our embedding. This choice represents a practical balance as choosing much larger values of $\lambda$ tends to produce an embedding graph that is increasingly fragmented and sparse, making the quantum evolution on the resulting graph trivial to simulate classically.

\begin{figure*}[!t]
\centerline{
\includegraphics[width=\textwidth]{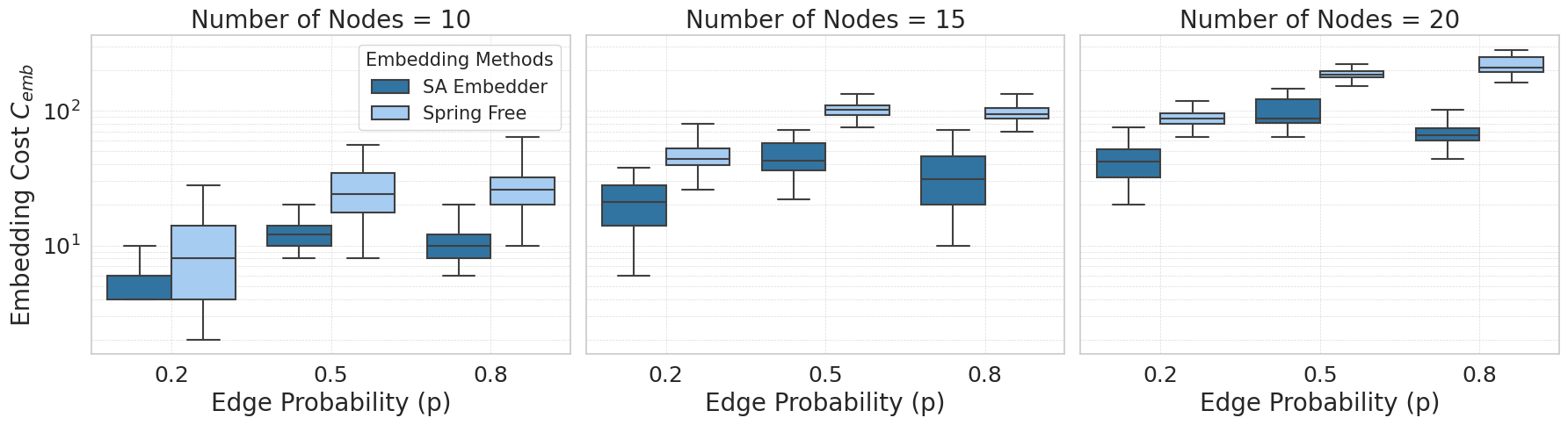}}
 \caption{Comparison of the \textsc{SA-Embedder} and "Spring Free" embedding methods, as described in the text. We compare the average final value of $C_2(G_H, G)$ (see Eq.~\eqref{eq:cost}) for 100 random Erdos-Renyi connected graphs $G$ with varying edge probability $p$. The three graphs correspond to varying sizes of the graph $G$, with respectively $10$, $15$ and $20$ nodes. The layout used for \textsc{SA-Embedder} is triangular. We see that, consistently, \textsc{SA-Embedder} recovers better embeddings which have less extra edges in the final graph $G_H$.}
 \label{fig:register}
\end{figure*}

In \textsc{SA-embedder}, we begin by identifying initial positions for the atoms using "Spring Free", which we then snap into the given layout graph $H$ (in Fig.~\ref{fig:SA-workflow} we used a triangular lattice but any layout could be used) by assigning each vertex to the closest available site in $L$. This creates $G_H(0)$, the initial embedding to refine. We then improve the embedding using a simulated annealing process. Iteration $k=0$ starts at an initial inverse temperature $\beta_{k=0} = \beta_i$ (typically, $\beta_i = 0.1$). At each iteration of the annealing, the inverse temperature is increased $\beta_k = \beta_{k-1}/\alpha$, with $\alpha$ the cooling factor. We choose $\alpha = 0.985$ in this implementation. It is expected that, as the graph size increases, the cooling factor will need to be increased so as to ensure thermalization at each step. At a given temperature $\beta_k$,  $N_{it}$ moves are attempted to reach thermalization - we choose $N_{it} = 150$ for the examples below. The algorithm stops when  $k=k_{\max}$ (equivalently, when $\beta_k = \beta_f \equiv \beta_i/\alpha^{k_{\max}}$)  or the objective value $\mathcal{C}$ has not improved in $dk$ steps. We choose $k_{\max} = 500$ and $dk = 40$.

Moves consist of two types of actions, where $G_H(k-1) \rightarrow G_H(k)$, each chosen with equal probability: (i) move, where an atom is relocated to an unoccupied site on the layout, and (ii) swap, where two atoms exchange positions. Nodes could be chosen equivalently, but we choose to introduce some bias by selecting randomly from the distribution of $\{p_i = c_{\lambda,i}/\mathcal{C}_{\lambda} | i \in V\}$ (using $\mathcal{C}_{\lambda} \equiv \mathcal{C}_{\lambda}(G_H(k-1), G)$) - this biases the choice towards nodes whose neighborhood was poorly embedded (high contribution to the loss function of Eq.~\eqref{eq:cost}). These moves then have an associate change in the loss function $\Delta \mathcal{C} = \mathcal{C}_{\lambda} (G_H(k), G) - \mathcal{C}_{\lambda} (G_H(k-1), G)$. Because moves are local, this calculation is fast and only involves lookup of a single row and column in the corresponding adjacency matrices. Acceptance of the proposed action follows the Metropolis-Hastings criterion: it is accepted if $p \in \text{Uni}(0,1)$, a random number, is found to be $p \leq \min(1, \exp (-\beta \Delta \mathcal{C}))$.

\begin{figure}[!h]
\centering
\includegraphics[width=\textwidth]{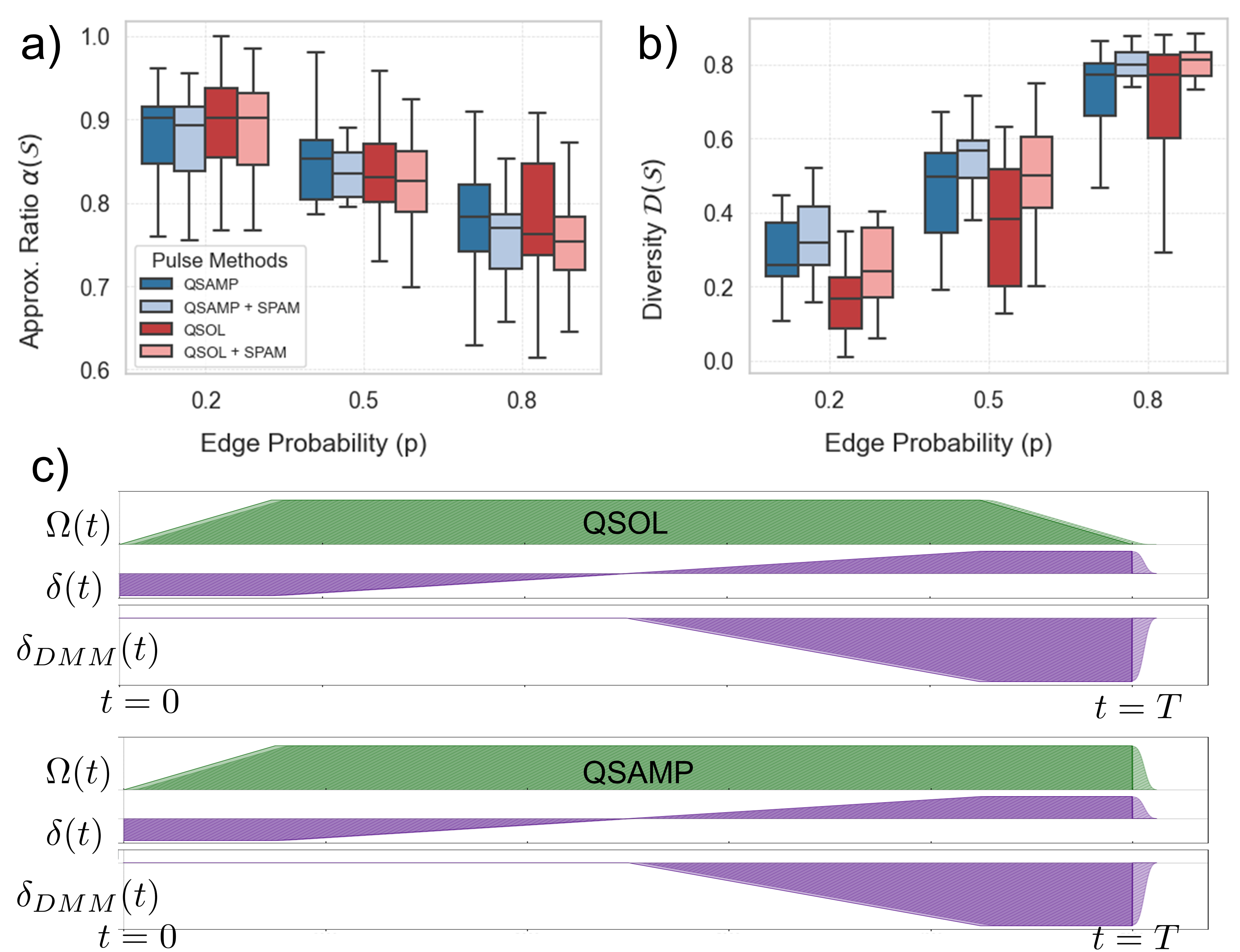}
\caption{Comparison of quality (a) and diversity (b) of the $M=1000$ measurement samples for the two pulse strategies (QSOL, see (c) top, and QSAMP, see (c) bottom). Quality is evaluated through the approximation ratio $\alpha(\mathcal{S})$, while diversity is evaluated through the bitwise difference metric of Eq.~\eqref{eq:diversity} Benchmarks are done on 100 random Erdos-Rényi graphs with $N=30$ nodes and varying edge probability $p$, with node weights $w_i$ drawn uniformly from $[1, 10]$. Any obtained bitstring is passed through the classical post-processing. We compare results with and without SPAM errors. While QSAMP delivers similar approximation ratio to QSOL across all edge probabilities, it delivers a more diverse set of samples for small $p$.}
\label{fig:pulse_benchmark}
\end{figure}

We compare \textsc{SA-embedder} with "Spring Free", a version of the Fruchterman-Reingold routine from Ref.~\cite{coelho_efficient_2022} with an added step. After a real space positioning of the atoms is found, an iterative search for the optimal Blockade radius (using the loss function $\mathcal{C}_\lambda$ with $\lambda = 2$) is performed. Note that, while \textsc{SA-Embedder} operates on a fixed layout, "Spring Free" places the atoms in real space. Benchmarks are performed on 100 connected Erdos-Rényi random graphs of varying sizes $N$ and edge probability $p$ (the number of edges will be $|E| = p N(N-1)/2$). Fig.~\ref{fig:register} shows that \textsc{SA-embedder} consistently achieves lower values of the loss function $\mathcal{C}_\lambda$ with $\lambda = 2$. We find improvements of up to an order of magnitude on larger and denser graphs, and thus our embeddings have lower extra edges than those obtained with "Spring Free". We see that, as graph sizes or density increases, embedding quality degrades; this is expected, as it becomes less likely the graph is a UDG. Furthermore, we see that, for graphs with higher edge probability, \textsc{SA-embedder} leads to substantially better values of $\mathcal{C}_2$ than "Spring Free" - the embedding has less extra edges. 

A brief comment on how embedding quality with \textsc{SA-embedder} may scale beyond the tested sizes is useful. 
Since the embedding problem is typically NP-hard for general graphs, the number of updates affecting $\mathcal{O}(1)$ atoms to reach the optimal value loss function value $C_{\lambda}$ is expected to scale exponentially with system size unless the graph has a trivial structure (e.g., a 1D chain, fully disconnected, or fully connected graphs). Moreover, the optimal value $C_{\lambda}$ is expected to grow as $N$ increases due to the increasing number of potential edges $\mathcal{O}(pN^2)$, where $p$ is the edge probability, that can cause UDG violations. Note that the dependence on $p$ is non-trivial: while increasing $p$ generally increase the optimal $C_{\lambda}$, the extreme case $p=1$ (complete graph) has a perfect UDG representation. To improve embedding quality for larger systems, one could consider: 1) an enhanced annealing schedule with an increased number of iterations, and 2) cluster updates in which $\mathcal{O}(N)$ atoms are moved or swapped simultaneously. In this context, cluster moves appear particularly promising for improving embedding quality, since in larger graphs, we found that the nodes naturally tend form separate disconnected clusters at high $\beta$, and connecting them would require moving entire clusters close to each other.

\subsection{Pulse Design} \label{sec:pulse}

The usual design for the $\Omega(t),\delta(t)$ waveforms is a form of quantum annealing. The core idea is to adiabatically evolve the system from an easy-to-prepare initial ground state $|\psi_0\rangle = |0\rangle^{\otimes N}$ to the ground state of a final Hamiltonian. By continuously varying the external controls slowly, the system remain in the instantaneous ground state \cite{Albash_2018} - this enables us to sample components from the low-lying energy states effectively. In the QSOL strategy, we proceed with the usual ramp-sweep-fall protocol \cite{henriet_quantum_2020, leclerc2024quantum}. This protocol is shown in Fig.~\ref{fig:pulse_benchmark} (c), and involves a three step process. From $t=0$ to $t=0.15T$, $\Omega(t)$ is linearly increased from $0$ to $\Omega_{\max}$, while the detuning is fixed to $\delta_i = - 2\Omega_{\max}$. From $t=0.15T$ to $t=0.85T$, $\Omega(t) = \Omega_{\max}$, while the site dependent detuning is swept from $-2\Omega_{\max}$ to $2\omega_i \Omega_{\max}$ ($\omega_i$ represents the weight on the nodes in the MWIS problem). This can be achieved using the DMM. Finally, from $t=0.85T$ to $t=T$, $\Omega(t)$ is brought continuously to $0$ while the detuning is stable.

We compare this typical pulse strategy with QSAMP, where we make a small alteration: in the final step of the QSOL protocol, instead of lowering the Rabi frequency, we keep it fixed to $\Omega_{\max}$ until measurements are performed. The objective of this small change is so that we promote quantum fluctuations and further explore computational states that are nearby in energy.




We benchmark the performance of these two pulse strategies in terms of quality and diversity across 20 different MWIS instances of weighted Erdos-Rényi random graphs with $|V| = 30$ nodes  and edge probability $p$. The weights of the nodes were drawn uniformly from the interval $[1, 10]$. Manipulation of the quantum pulses is done using the \textsc{Pulser}~\cite{silverio_pulser_2022} Python packages. Emulations of $N\leq 12$ qubits are done using the QuTip backend~\cite{johansson2012qutip}, while those with $N>12$ are done using \textsc{emu-mps}~\cite{bidzhiev2023cloud}, an open-access tool for matrix-product state (MPS) simulations of neutral-atoms quantum systems. The emulator's parameters are kept in their default settings.

Consider a set of $M$ samples $\mathcal{S} = \{ \Vec{s}_1, \Vec{s}_2, \dots, \Vec{s}_M \}$. Our metric for quality is the approximation ratio $\alpha(\mathcal{S}) = \sum_{m = 1}^{M} c_G(\Vec{s}_m)/c_G^{\text{opt}}$, which we average over graph instances $G$, and where $c_G(\Vec{s}) = \sum_i s_i \omega_i$ is the cost of the independent set $\vec{s}$ in the weighted graph $G$ and $c_G^{\text{opt}}$ is the optimal value determined using an exact solver. Our metric for diversity is obtained through the Hamming distance of samples, which can be written as
\begin{equation}
    \mathcal{D}(\mathcal{S})= \frac{1}{2\bar{s}\binom{M}{2}} \sum_{1 \leq n \leq m \leq M} \lvert s_n \oplus s_m \rvert \label{eq:diversity} 
\end{equation}
where $|\cdots|$ denotes the size of a given set, $\bar{s} = \frac{1}{M}\sum_{m=1}^{M} |\Vec{s}_m|$ is the mean size of the independent sets,  and $\oplus$ is the exclusive OR operator, which takes the symmetric difference between sample $s_n$ and $s_m$. The denominator ensures a normalization to $1$, and again we average this quantity over graph instances $G$.

We use our \textsc{SA-embedder} method and evaluate the performance of the two pulses on the same register, with $M=100$ samples. Each sample undergoes the "Maximalize" postprocessing described in Appendix~\ref{sec:Appendix-Greedy}. We also evaluate the performance of the two pulse strategies under state preparation and measurement (SPAM) noises that are inherent to any QPU. The parameters of the SPAM are described in Table III of Ref. \cite{coelho_efficient_2022}.
Fig \ref{fig:pulse_benchmark} summarizes the results. The quality of solutions is similar across both pulse strategies and is robust to SPAM noise. This is largely due to the effectiveness of the greedy postprocessing, which can compensate for local bit flips. In contrast, diversity is significantly higher for QSAMP, and SPAM noise appears to slightly enhance diversity. Moreover, solution quality (diversity) decreases (increases) as the edge probability increases. This is linked to the limitations of the embedding strategy: as the graph becomes dense, the quality of the UDG approximation of this graph becomes significantly worse and there are more missing edges. In this regime, the solution quality is mostly due to the greedy postprocessing step, having been hot-started by the quantum pulses. 



\begin{figure*}[t]
  \centering
  \includegraphics[width = 0.8\textwidth]{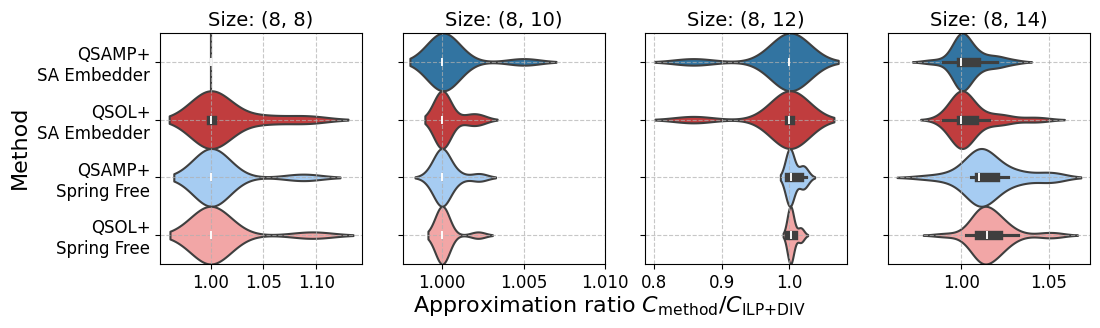}
  \includegraphics[width = 0.8\textwidth]{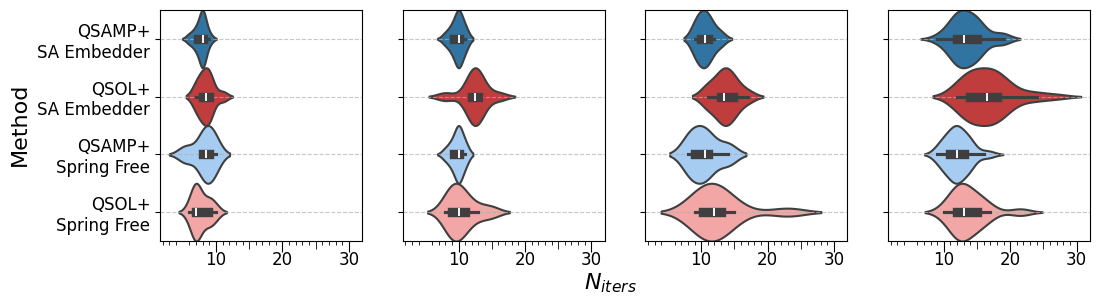}
  \caption{Comparison of pulse and register design protocols for the quantum methods tackling the PSP in the CG workflow for the fleet assignment problem. Each column corresponds to an instance class $(|V|, |K_v|)$ of the fleet assignment problem, as described in Subsection~\ref{subsec:instance}. (Top): approximation ratio $C_{\text{method}}/C_{\text{ILP+DIV}}$, where we see that, as the instance size $|K_v|$ increases, methods using \textsc{SA-embedder} lead to lower (better) approximation ratios. (Bottom): Average number of iterations $N_{\text{iters}}$ for the CG workflow. We see that QSAMP pulse designs lead to the smallest number of iterations (i.e. faster termination of the workflow).  
  }
  \label{fig:OBJ_VAL_quantum}
\end{figure*}

\begin{figure*}[t]
  \centering
  \includegraphics[width = 0.8\textwidth]{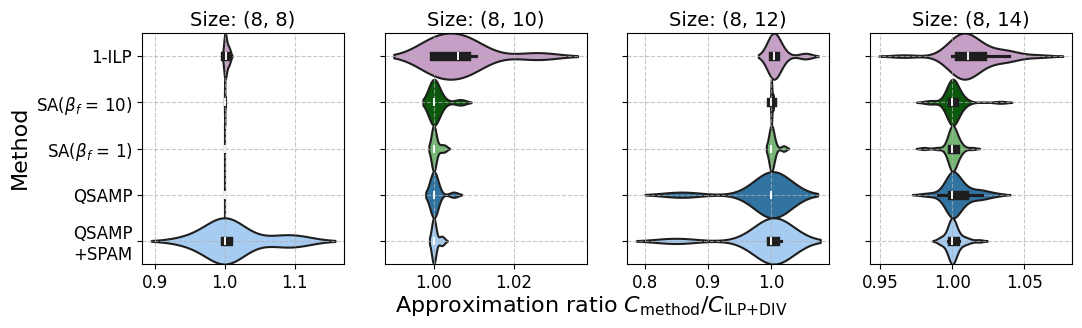}
  \includegraphics[width = 0.8\textwidth]{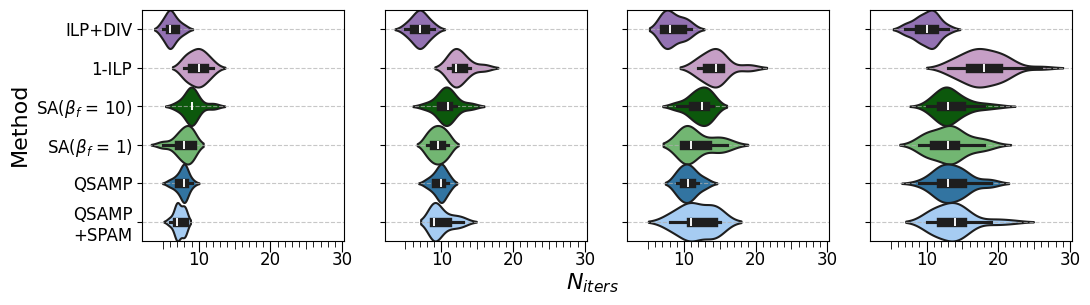}
  \caption{Comparison of our proposed quantum protocol, QSAMP with \textsc{SA-embedder}, with and without SPAM errors, with the classical methods described in Sec.~\ref{sec:classical}. Each column corresponds to an instance class $(|V|, |K_v|)$ of the fleet assignment problem, as described in Subsection~\ref{subsec:instance}. (Top): approximation ratio $C_{\text{method}}/C_{\text{ILP+DIV}}$, where we see quantum methods that are generally on-par with classical SA-based methods. 1-ILP falls short due to the lack of column intensification. (Bottom): Average number of iterations $N_{\text{iters}}$ for the CG workflow. We see that ILP+DIV leads to the least amount of iterations, while QSAMP and SA($\beta_f = 1$) are on par with each other.}
  \label{fig:OBJ_VAL_benchmarks}
\end{figure*}

\section{Results} \label{sec:results}

As previously mentioned, the complexity in CG schemes is moved to the sampling of high quality solutions to the PSP. In this section, we benchmark the performance of the classical methods presented in Sec.~\ref{sec:classical} and the quantum methods presented in Sec.~\ref{sec:analog}. These results are obtained on synthetic instances of the fleet assignment problem, whose construction is described in Subsection~\ref{subsec:instance}.

We found early on that the solver that returned the best results most often was ILP+DIV. This is logical, as this solver was constructed with the specific intent of returning the $M$ best diversified columns, thus maximizing the objectives of column diversification, intensification and quality. We also compared our results with the Gurobi~\cite{gurobi} black-box solver. While Gurobi sometimes found better results than our CG scheme, it also sometimes stayed stuck in local minima and resulted in worse solutions. We thus chose to compare all methods to ILP+DIV, although we add comparisons to Gurobi in Appendix~\ref{app:gurobi}, as well as details on its implementation. Similarly, we found that the "Greedy" method gave such subpar results compared to all other methods that it skewed data visualization. Although we tested it on all instances, we opted not to show it in the graphs below.

In Subsection~\ref{subsec:compare_quantum}, we compare the performance on the fleet assignment problem of our quantum routines QSOL and QSAMP while using different embedding methods. We find that QSAMP with \textsc{SA-Embedder} consistently returns better objective values in less iterations. Then, in Subsection~\ref{subsec:benchmark_classical}, we compare this quantum routine with the classical alternatives, showing that classical SA samplers are most performant for our instances. We then remark that adding a simple step called Make\_Diff (described in Appendix~\ref{sec:Appendix-Greedy}) to our postprocessing leads to major improvements for the quantum routine. Finally, in Subsection~\ref{sec:quality}, we explicitly compare the quality and diversity objectives of the PSP solvers.


\subsection{Synthetic instances} \label{subsec:instance}

At each iteration of the CG scheme, there are $|V|$ PSPs that must be solved over subgraphs $G_v$ for all $v \in V$. To ensure a control over the hardness of our instances, we make sure that all $G_v$ remain hard (i.e. non-trivial).
We control the number of vertices $\abs{K_v}$ (i.e. the number of compatible tours per vehicle class), the number of edges $\abs{E_v}$ via the edge's probability $p = \abs{E_v}/\binom{\abs{K_v}}{2}$ (i.e. the number of conflicting tours). 
We also control the number of vehicle classes $\abs{V}$ (i.e. the number of independent PSP${}_v$ to solve at each iteration). There are also parameters influencing the complexity of the RMP: the average number of allowed vehicle types per tours $\bar{N_V} = 1/\abs{K} \sum_{k=1}^{\abs{K}} |V_k|$, from which we infer the total number of tours in the problem as $K = \lceil \bar{N_V} \abs{K} \rceil $, the costs of each tour $C_l^{(T)}$ and vehicle types $C_v^{(v)}$, and the availability bounds $N_{v}^{\min}$ and $N_v^{\max}$ for each vehicle type $v$. Our synthetic instances have the following global parameters: $\abs{V} = 8$, graphs $G_v$ are Erdos-Renyi connected graphs with $p \sim 0.30$, $\bar{N_V} = 2$, $C_k^{(T)} \in \mathcal{N}(10, 5)$, $C_v^{(v)} \in \mathcal{N}(50, 10)$, $N_v^{\min} = 1$ and $N_v^{\max}$ where $\mathcal{N}(\mu, \sigma^2)$ is a normal distribution with mean $\mu$ and variance $\sigma^2$. The  Instance classes are noted by the tuple $(|V|, |K_v|)$. We create 10 random instance (random generation of subgraphs and costs following the above distributions), for the $(8,8)$, $(8,10)$ and $(8,12)$ instances. For the $(8,14)$ instances, we generate 30 instances in order to perform a more in-depth analysis of the impact of the chosen solver's performance on the CG scheme.

Note that, in all results presented in this section, all classical and quantum methods, except 1-ILP, are asked to return $M=5$ samples (potential columns) per PSP${}_v$ subproblems per CG iteration. Thus, up to $5|V|$ columns could be added at each iteration, although less may pass the reduced cost threshold for acceptance.

\subsection{Comparison between quantum methods} \label{subsec:compare_quantum}

In Fig.~\ref{fig:OBJ_VAL_quantum}, we compare the proposed register and pulse design protocols on the CG task for the fleet assignment problem, accross the four instance classes described in Subsection~\ref{subsec:instance}. Results are shown in the "violinplot" format (from the \textsc{Seaborn} Python library). These plots feature a kernel density estimation of the underlying distribution~\footnote{Note that the kernel density estimation procedure is influenced by the sample size, and violins for relatively small samples might look misleadingly smooth. This may be the case for $(|V|, |K_v|)$ instances with $|K_v| = 8,10,12$, where only $10$ instances are compared.}. White lines refer to the median of the data, while the black box represents the interquartile range (range that excludes the bottom and top $25\%$ of data).

In the first row, we show the approximation ratio $C_{\text{method}}/C_{\text{ILP+DIV}}$, where $C_{\text{method}}$ is the final cost of the solution to the fleet assignment problem found using the CG workflow with "method" as the PSP solver. On average, all methods thus return a final objective as good as ILP+DIV for the $|K_v| = 8,10,12$ instances. As the instance size is increased to $|K_v| = 14$ (thus, the graphs for the sub-PSPs become larger), then a clear distinction appears: methods using \textsc{SA-embedder} lead to better results than those using "Spring Free". This is in line with the results shown in Fig.~\ref{fig:register}, where "Spring Free" is shown to lead to poorer embeddings with a larger number of extra edges. These edges prevent the exploration of valid independent sets on the graph, and thus column quality is lower, which impedes the overall performance of the CG scheme. In fact, a poor embedding strategy leads to an increasingly important role of the greedy post-processing compared to the quantum sampling, and greedy solvers perform poorly for these PSP subproblems. 

In the second row, we wee that, as the system size $|K_v|$ of the sub-PSP is increased, stark differences between the QSAMP method and the QSOL method appear, with the latter taking generally less iterations to reach the termination condition. As QSAMP returns more diverse columns at each iterations, it may accept at early iterations columns that QSOL may take many iterations to find as nearly optimal ones. Therefore, QSAMP can do more in less iterations by preemptively generating many good diverse columns.

\subsection{Benchmark with classical alternatives} \label{subsec:benchmark_classical}

In light of the results presented in Fig.~\ref{fig:OBJ_VAL_quantum}, we select the QSAMP method with \textsc{SA-embedder} as our leading quantum method for comparison with classical counterparts. These results are presented in Fig.~\ref{fig:OBJ_VAL_benchmarks}, where we compare its performance with ILP+DIV (used as the reference method), 1-ILP, SA($\beta_f = 10$), SA($\beta_f = 1$) methods as described in Subsection.~\ref{sec:classical}. We also tested a version of the QSAMP quantum protocol with SPAM errors (the same parameters as in Sec.~\ref{sec:pulse} are used). It is found that, while SPAM errors lead to fluctuations in the generated columns, the median result stayed similar and the overall performance seemed unaffected - an encouraging sign for the deployment of this workflow in the NISQ era.

Again, in the top row, we show the approximation ratio $C_{\text{method}}/C_{\text{ILP+DIV}}$, while the bottom row shows the number of iterations $N_{\text{iters}}$ taken until termination of the CG workflow. First, we see the clear effects of column intensification as all solvers outperform 1-ILP (which adds at most one column per iteration) across all problem sizes. This subpar performance is in spite of a larger number of iterations. This is consistent with the literature (Ref.~\cite{desauliers2005column}) and the results of \cite{da_silva_coelho_quantum_2023}. On the other hand, ILP+DIV takes the least number of iterations, owing to its ability to add many of the best columns to the RMP at each iteration.

Then, we see that, in the classical SA-based samplers (in green), the final objective value reached is generally unchanged whether one uses $\beta_f = 1$ or $\beta_f = 10$, while using $\beta_f = 1$ leads to a smaller number of iterations $N_{\text{iters}}$. The quantum protocol QSAMP led to similar results as the classical SA-based samplers, both in the final objective and in the number of iterations.

Analyzing in detail these results, we  realized that the samples coming out of the QSAMP method were still sometimes degenerate - when asked to return $5$ samples, $2$ or $3$ might be the same independent set. This situation never occurred for ILP+DIV, and very infrequently for SA($\beta = 1$). Thus, we developed another layer of greedy postprocessing which seeks to ensure that all samples generated by a given sampler are non-degenerate. It does so by randomly removing nodes from a degenerate IS until the new IS returned is not degenerate with the rest of the samples. A description of this algorithm is provided in Appendix \ref{sec:Appendix-Greedy}.

We present in Fig.~\ref{fig:3Algos} confusion plots which compare, instance by instance for the $30$ instances of the class $(8,14)$, the performance of different methods in either return better, equal or worse objective value $C_{\text{obj}}$, or in taking less, equal or more iterations $N_{\text{iters}}$. For the bare results, we see that QSAMP generally takes more iterations than SA($\beta_f = 1$) for a similar objective (first subplot, top row), while both QSAMP (second subplot, top row) and SA($\beta_f = 1$) (third subplot, top row) take more iterations that ILP+DIV for a worse result. When Make\_Diff is applied to all methods, the situation changes, and QSAMP now takes less iterations than SA($\beta_f = 1$) while its final objective value is overall better (first subplot, bottom row). We also see that QSAMP is now able to have on par performance with ILP+DIV (second subplot, bottom row), while taking more iterations. SA($\beta_f = 1$) is also improved with respect to ILP+DIV (third subplot, bottom row), but less so than QSAMP. This shows that the Make\_Diff tool is most useful for the quantum samples from QSAMP, confirming that the samples from SA($\beta_f = 1$) were already sufficiently non-degenerate. This cheap postprocessing is thus able to recover results for our quantum pricing heuristic that are on par with ILP+DIV, a costly exact method.

\begin{figure}[!t]
\includegraphics[width=0.7\textwidth]{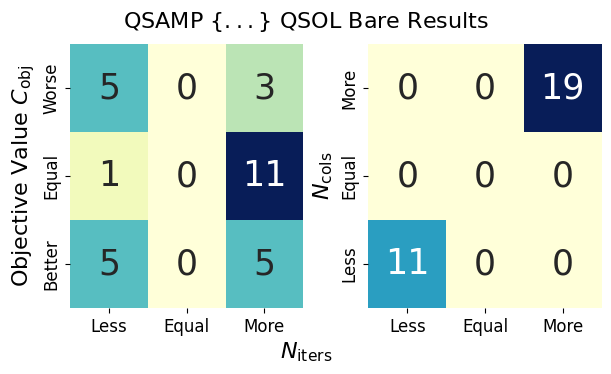}
\includegraphics[width=0.7\textwidth]{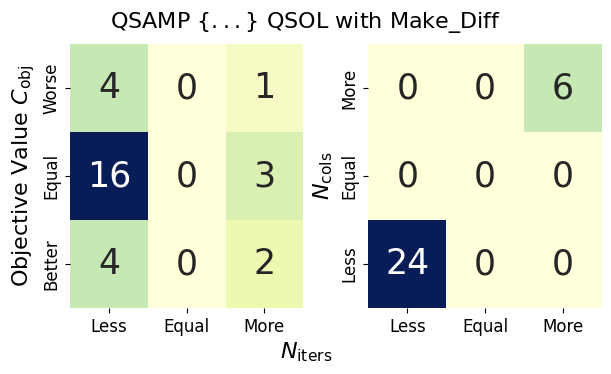}
 \caption{Confusion Plots comparing QSOL and QSAMP quantum PSP routines, for $30$ instances of the problem class $(8,14)$. The top row corresponds to the bare results, while the bottom row represents the results with the addition of the Make\_Diff subroutine. Left column compares the objective value ($y$ axis) with the number of iterations ($x$ axis), while the right column compares the number of accepted columns to the RMP ($y$ axis) with the number of iterations ($x$ axis). With Make\_Diff, QSAMP and QSOL methods return final solutions with comparable quality, although this comes at a cost: QSOL takes more iterations and generates more columns than QSAMP.}
 \label{fig:QSAMP-QSOL}
\end{figure}

We compare the impact of the Make\_Diff postprocessing on the QSAMP and QSOL methods in Fig.~\ref{fig:QSAMP-QSOL}. We see the drastic impact of it on the samples obtained. The bare results show that both methods generally lead to similar final objective value, the number of iterations taken and the total number of columns generated $N_{\text{cols}}$ seems stochastic ie sometimes being less and sometimes being more. The Make\_Diff postprocessing bias things such that, for the same objective value, the QSAMP method takes less iterations and generates less columns. This reinforces the importance for a sampler that generates many, good, and diverse columns. Samplers with these parameters lead to good final objective values that terminate in less iterations and require less columns. For NISQ implementations of this heuristic on the fleet assignment problem and other CO problems that allow a CG decomposition, this is good news, as each sample corresponds to a call to the quantum computer. Quantum protocols that promote both high quality and high diversity samples are thus more efficient.

\begin{figure}[!t]
\includegraphics[width=\textwidth]{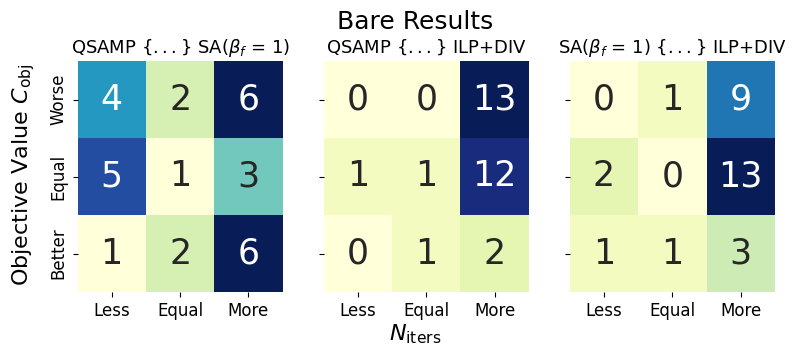}
\includegraphics[width=\textwidth]{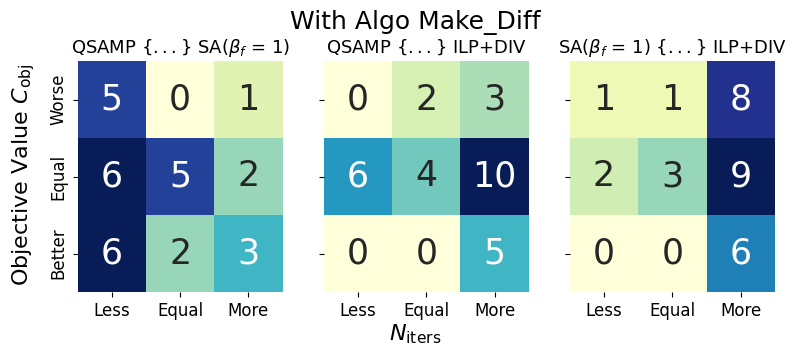}
 \caption{Confusion plots comparing pairs of PSP samplers. The title of each subplot refers to the pair of compared methods. For $30$ random instances of the problem class $(8,14)$, we compare the CG workflow in either returning lower values of $C_{\text{obj}}$, or taking less iterations. For a given square in a subplot, the $y$ or $x$ labels should be inserted in the title statement "Method1 $\{\cdots\}$ Method2", while the number indicates how many instances follow this statement. The $y$ axis statements are about the final objective value, while the $x$ axis statements are about the total number of iterations required. The top row represents the bare results, while in the second row we present results with the added Make\_Diff (see Appendix~\ref{algo:ILP+DIV}) routine that ensures non-degeneracy of the samples.  
 }
 \label{fig:3Algos}
\end{figure}

\begin{figure}[!t]
\includegraphics[width=0.85\textwidth]{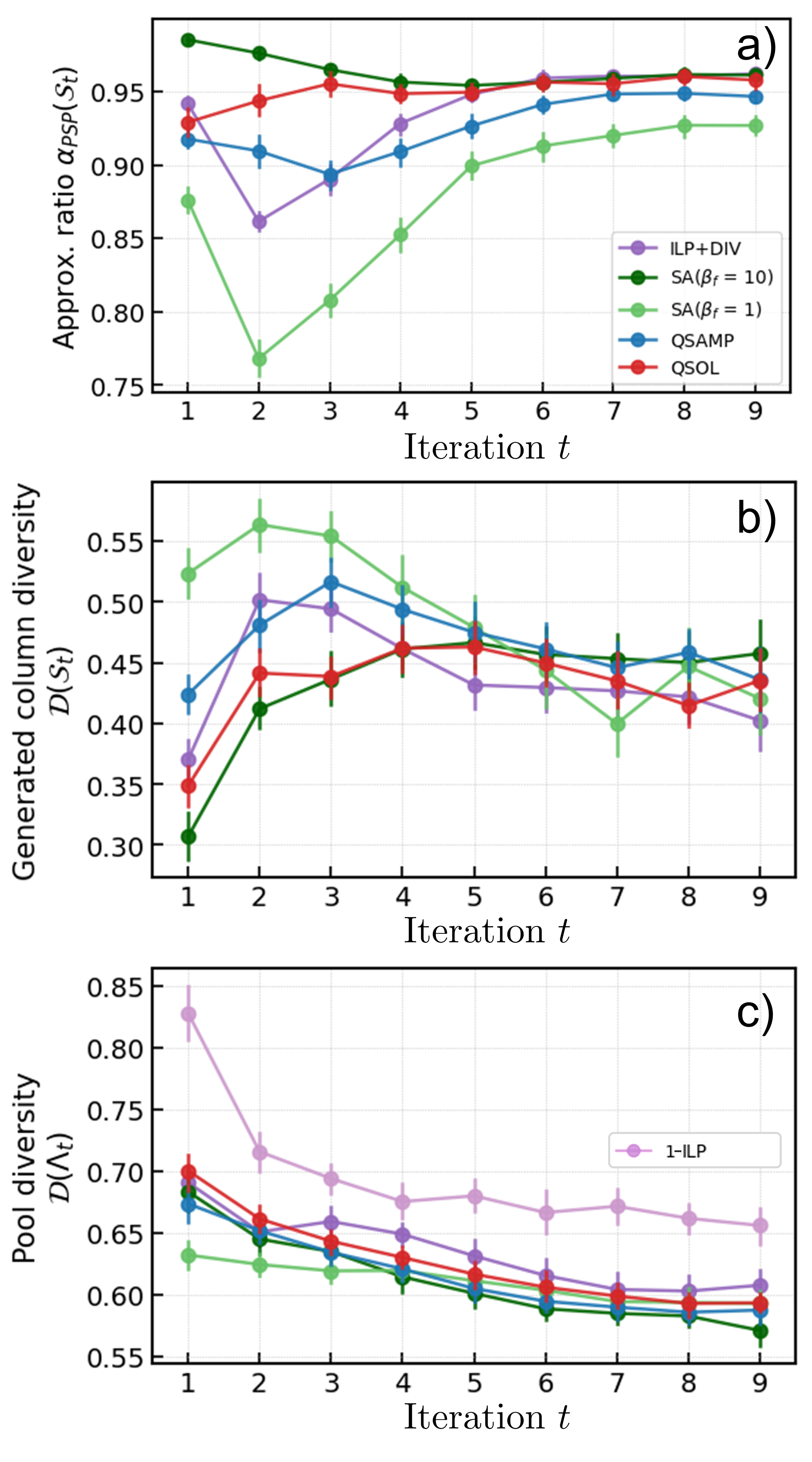}
\caption{Evolution of the approximation ratio $\alpha_{\text{PSP}}(\mathcal{S}_t)$ (a), the diversity of the generated columns $\mathcal{D}(\mathcal{S}_t)$ (b) and the diversity versus the current pool of columns $\mathcal{D}(\Lambda_t)$ (c). The data points are averages over the 30 instances of $(\abs{V}, \abs{K_v}) = (8, 14)$. At each CG iteration $t$, the generated column set is $\mathcal{S}_t$, while the set of columns currently in the RMP is $\Lambda_t = \Lambda_{t-1} \bigcup \mathcal{S}_{t}$. The measure for the diversity is in Eq.~\eqref{eq:diversity}. The Make\_Diff postprocessing method is used on all methods. 
}
 \label{fig:Iters}
\end{figure}

\subsection{Quality and Diversity Analysis} \label{sec:quality}



Figure \ref{fig:Iters} provides a per-iteration analysis of the quality and diversity of the set  $\mathcal{S}_t$ of columns generated by the PSP samples. Each point represents an average over the runs for the 30 instances of problem size $(8, 14)$. Data is obtained using the Make\_Diff postprocessing. The quality and diversity metrics are those introduced in Sec.~\ref{sec:pulse} (see Eq.~\eqref{eq:diversity}). In Fig.~\ref{fig:Iters} (a), we compare the average approximation ratio $\alpha_{PSP}(\mathcal{S}_t)$ for the accepted columns of the sub-PSPs at each iteration $t$. In (b), we show the diversity of the column set $\mathcal{D}(\mathcal{S}_t)$, and in (c) we show the diversity of the overall columns $\Lambda_t$ generated thus far in the CG scheme. 

The exact method 1-ILP, which returns the best column to add only, is not displayed in (a-b), as its approximation ratio is $1$ and its diversity is ill-defined. However, we see from Fig.~\ref{fig:Iters} (c) that this method consistently has the largest pool diversity. This means that the best column to add is also the one that is the most diverse with respect to the pool. All other methods shown in (c) follow the same pattern for the pool diversity - their differences shine in the other metrics.

We see that SA($\beta_f = 10$) returns very high quality columns, especially at early iterations, but it ranks among the worst with respect to diversity of the column set. On the other hand, SA($\beta_f = 1$) leads to a poor approximation ratio in the beginning of the CG scheme, while its diversity is very high. Both methods performed similarly in overall performance, as shown in Fig.~\ref{fig:OBJ_VAL_benchmarks}. We find that ILP+DIV, which  was the classical method that performed best in our CG scheme, hits the right balance of a high approximation ratio and a high diversity. We thus link this performance of ILP+DIV as a PSP sampler to its overall performance in the CG scheme. 

The quantum methods, on the other hand, show behavior expected from the initial benchmarks from Fig.~\ref{fig:pulse_benchmark}: QSOL has higher quality of solutions, while QSAMP has a higher diversity, especially at early iterations. We furthermore observe that QSAMP more closely mimics the behavior observed for ILP+DIV at early iterations, while QSOL tracks it more closely at later iterations. It is possible that a different SA method with a fine-tuned $\beta_f$ might also mimic the ILP+DIV behavior. 

In all methods, we see a transition from a period of diversification (generating columns of high diversity) to one of intensification (generating columns of high quality), occurring on average at around iteration 3. We thus conjecture that the QSAMP protocol is more efficient at early iterations in hybrid CG schemes, leading to high diversity samples, while the QSOL protocol is preferred for later iterations. It is however unclear how one can define a metric to efficiently transition from one to the other and improve the CG scheme's performance.

\section{Conclusion}

We showed in this paper that novel register embedding and pulse design protocols for analog neutral-atoms quantum computers can enhance the performance of hybrid column generation schemes. We showed that these protocols can compete with other classical solvers for the pricing subproblems in returning a large set of high quality and diversity columns at each iteration. We also introduced a new greedy postprocessing method called Make\_Diff which seeks to modify degenerate samples in order to obtain a set of potential columns that are all non-degenerate. Importantly, our quantum protocol with this postprocessing most closely mimics the behavior of an idealized solver that finds the $M$ best columns to add to the reduced master problem. 


The focus of this paper was on the fleet assignment problem, which acted as a hard combinatorial problem on which to test our hypotheses concerning the design and performance of analog quantum primitives in column generations schemes. We note that the conclusions and the methods presented here can be applied to other practical settings. In machine job scheduling, for example, jobs must be assigned to machines in a way that respects machine capabilities and job durations, aiming to optimize various cost metrics. Likewise, in airline and public transit scheduling, aircraft, or buses must be assigned to flight or route sequences while accounting for various constraints. 
In such industrial settings, one may need to expand the workflow to include branching (as was done in Ref.~\cite{vercellino2025hybrid}) or the generation of cuts \cite{desaulniers_cutting_2011}.

These settings all represent prime places on which an industrial quantum advantage may be sought. To achieve such an advantage, our paper places a threshold to meet: the quantum protocol must be able to return a set of $M$ high quality and high diversity columns at each iterations faster than competing classical methods, such as simulated annealing and tensor networks. 
This remains an open question, although the experimental results of quantum annealing protocols to solve large-scale maximum independent set problems on neutral-atoms hardware \cite{ebadi2022quantum, schuetz2025quantum, cazals2025quantum} are encouraging. 
 
We conclude by noting from Fig.~\ref{fig:OBJ_VAL_benchmarks} that the quantum protocol for the $(8,14)$ instances would take on average $\sim 14$ iterations, each using $5$ shots from the QPU while implementing a register of $14$ atoms. Assuming an available NAQC capable of returning 1 shot per second and hosting $100$ qubits \cite{PasqalOrion}, we can then estimate the runtime as $\sim 140$ seconds (we can run multiple graphs using the same global schedule, as long as they are well-separated on the QPU). On the other hand, our comparisons with Gurobi (see Appendix~\ref{app:gurobi}) show that even when given a time limit of $8$ hours, the solver returned with nearly equal likelihood better and worse solutions for our synthetic instances. This shows the potential of the hybrid CG schemes presented here as powerful heuristics.







\acknowledgements

We thank Wesley Coelho, Yassine Naghmouchi and Hossein Sadeghi for valuable discussions and support. This research was financially supported by the Natural Sciences and Engineering Research Council of Canada (NSERC) and Pasqal Canada through an Alliance grant (ALLRP 590810 - 23). This work made use of compute resources by Calcul Québec and the Digital Research Alliance of Canada.

\appendix
\section{ILP+DIV}
\label{sec:Appendix-ILP+DIV}
We present the ILP+DIV algorithm designed to generate a set of $M$ high-weight independent sets of a graph $G = (K, E)$ (with weights $w_i \in  \mathbb{R}$, $i \in K$) while ensuring non-degeneracy of the $M$ independent sets. Algorithm~\ref{algo:ILP+DIV} summarizes the procedure. The algorithm starts by initializing a list of constraints $\mathcal{E}$, which will be used to cut previously obtained solutions from being obtained when solving the MWIS, thus enforcing that a new solution with maximum weight is found. One can exclude any independent set $\mathcal{I}$ from being returned by the ILP solver by adding the following constraint:  $\sum_{k \in K} B_I[k] \cdot x_k \leq \abs{I} - 1$ where $B_I[k] = 1$ for indices $k \in \mathcal{I}$ (i.e. it is a binary vector for the IS). By iteratively solving the MWIS with a list of constraints that corresponds to the exclusion of the previously obtained solution. The set of obtained solution is guaranteed to be non-degenerate. For positive weights only, we are also guaranteed to find the $M$ best IS. In the presence of negative weights, this may not be the case (we may be missing some solutions due to the addition of a negatively weighted node). Specifically, suppose that the first solution generated is the independent set $\{1, 2, 3\}$, which corresponds to the maximum weighted independent set of the graph. However, a closely related solution -- $\{1, 2, 3, 4\}$ where $w_4 < 0$ may represent the second highest weighted independent set, which would be excluded by our exclusion constraint $x_1 + x_2 + x_3 \leq 2$.
\begin{algorithm}
\caption{ILP+DIV}
\label{algo:ILP+DIV}

\KwIn{Graph $G = (K, E)$ with node weights $w: K \to \mathbb{R}$; $M$ rounds}
\KwOut{List $\mathcal{I}$ of independent sets and corresponding weights $\mathcal{C} = \{c_a \: | \: a \in \mathcal{I}\}$}

Define the modified ILP solver \texttt{ILP}($G, w, \mathcal{E}$)

\Fn{\texttt{ILP}($G, w, \mathcal{E}$)}
{
    Define $x_k \in \{ 0,1 \}$ for all $k \in K$\;
    Add constraints: $x_k + x_{k'} \leq 1$ for all $(k,k') \in E$\;
    \ForEach{$(B_S, \abs{S}) \in \mathcal{E}$}{
        Add constraint: $\sum_{k \in K} B_S[k] \cdot x_k \leq \abs{S} - 1$ 
    }
    Set objective: $\max \sum_{k \in K} w_k \cdot x_k$\;
    Get exact solution to ILP:  $x_k^{\text{opt}} \; \forall \: k \in K$\;
    $I \gets \{k \in K \mid x_k^{\text{opt}} = 1\}$\;
    \KwOut{Maximum Independent Set $I$ of $G$ under constraints $\mathcal{E}$}
}

Initialize $\mathcal{I} \gets [\,]$, $\mathcal{C} \gets [\,]$\;
Initialize constraint list $\mathcal{E} \gets \{\}$\;
Let $r \gets 0$\;
\While{$r < M$}{

    $I \gets \texttt{ILP}(G, w, \mathcal{E})$\;
    \If{$|I| = 0$}{
        \textbf{break}\;
    }
    $c_I \gets \sum_{k \in I} w_k$\;

    $\mathcal{I} \gets I$\;
    $\mathcal{C} \gets c_I$\;
    \tcp{Exclude this solution from future rounds}
    Set $B_I = \{0 \: \forall k \in K\}$\;
    \ForEach{ $k \in I$}{$B_I[k] \gets 1$}
    $\mathcal{E} \gets E \; \bigcup \;  (B_I, \abs{I})$\;
    $r \gets r + 1$\;
}

\Return{$(\mathcal{I}, \mathcal{C})$}

\end{algorithm}



    




\section{Greedy Post-Processing} \label{sec:Appendix-Greedy}

We present a greedy post-processing algorithm designed to validate and maximalize independent sets generated by  samplers that are likely to faulty independent sets. Algorithm \ref{algo:maximalize} summarizes our approach. The procedure first corrects any violations of the independent set constraint by removing the lowest-weight conflicting vertices. It then improves each independent set by greedily adding the highest-weight vertices that are "free", i.e. none of their neighbors are selected in the current independent set. This algorithm has a runtime that scales like $O(|E| + |V|)$ per independent set.

This post-processing can be coupled with another method: Make\_Diff, which we introduce in the main text. It's detailed implementation is shown in Algorithm \ref{algo:make_diff}. This algorithm ensures the uniqueness of the independent sets generated. It detects duplicates and iteratively removes the lowest-weight vertices from duplicate sets and create a reduced graph where this vertex is removed. It then re-applies the maximalization algorithm of Algo. \ref{algo:maximalize} on this reduced graph. This algorithm has a runtime of $O(\bar{M}^2 + \bar{M} |V|(|E| + |V|))$ in the worst case, where $\bar{M} \leq M$ is the number of non-degenerate samples. This algorithm is fast for a small number of samples, which is the case in this paper ($M=5$). In fact, for $M \ll V$, then it is dominated by the cost of the maximalize routine.
\begin{algorithm}
\caption{Maximalize}
\label{algo:maximalize}

\KwIn{Graph $G = (K, E)$ with weights $w: V \to \mathbb{R}$, and a list $\mathcal{I}$ of $M$ samples}
\KwOut{List $\mathcal{I}_{\text{clean}}$ of $M$ maximal independent sets}

Initialize $\mathcal{I}_{\text{clean}} \gets [\,]$\;

\ForEach{$I \in \mathcal{I}$}{
    
    \tcp{Fix conflicts}
    \While{$I$ contains adjacent nodes}{
        Remove conflicting node $k'$ in $I$ with smallest weight $w_{k'}$\;
    }

    \tcp{Maximalize}
    Let $N \gets$ all neighbors of nodes in $I$\;
    Let $C \gets K \setminus (I \cup N)$ \tcp*{Candidate nodes}
    Sort $C$ by decreasing $w$\;
    
    \ForEach{$v \in C$}{
        \If{$v$ not adjacent to any node in $I$}{
            $I \gets I \cup \{v\}$\;
        }
    }

    $\mathcal{I}_{\text{clean}} \gets \mathcal{I}_{\text{clean}} \; \bigcup \; \{I\} $\;
}

\Return{$\mathcal{I}_{\text{clean}}$}
\end{algorithm}

\begin{algorithm}
\caption{Make\_Diff}
\label{algo:make_diff}

\KwIn{Graph $G = (K, E)$ with weights $w: V \to \mathbb{R}$, and a list $\mathcal{I}$ of sampled independent sets}
\KwOut{List $\mathcal{I}'$ of unique independent sets}


Initialize output list $\mathcal{I}' \gets [\,]$\;

\ForEach{$I \in \mathcal{I}$}{
    Let $I' \gets I$\;
    \If{$I' \in \mathcal{I}'$}{
        Sort nodes in $I'$ by increasing weights $w_{\text{inc}}$: $[w_{k_0}, w_{k_1}, \ldots]$\;
        Initialize $\text{drop\_idx} \gets 0$\;
        \While{$I' \in \mathcal{I}'$ \textbf{and} $\text{drop\_idx} < |I|$}{
            Choose weight $w_{k_{\text{drop\_idx}}}$ and remove it from $I'$.\;
            \tcp{Maximalize algorithm}
            $I' \gets \texttt{Maximalize}(G', w, I')$ \;
            $\text{drop\_idx} \gets \text{drop\_idx} + 1$\;
        }
    }

    $\mathcal{I}' \gets \mathcal{I}' \; \bigcup \; \{I'\} $\;
}

\Return{$\mathcal{I}'$}
\end{algorithm}

\begin{figure*}[t!]
    \centering
    \includegraphics[width=1.0\linewidth]{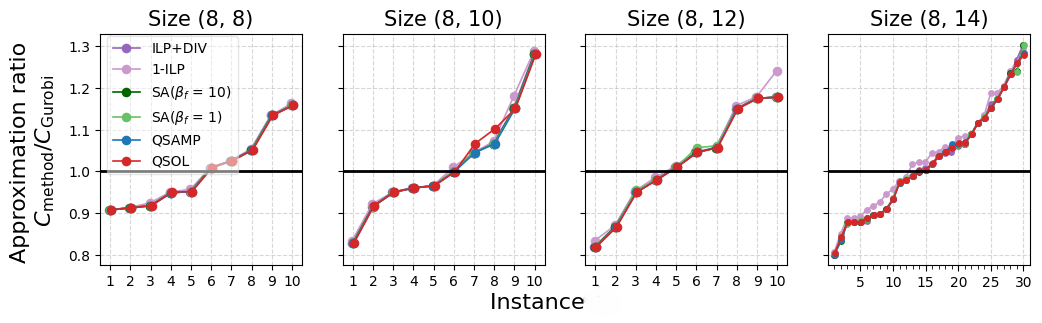}
    \caption{Instance-wise comparison of our various methods for the sampling of solutions to the PSP in our CG workflow against Gurobi. Each Gurobi evaluation is limited to a maximum of 8 hours and executed on 5 CPU cores (Intel 6972P @ 2.4 GHz). Our CG workflow found better solutions than Gurobi ($C_{\text{method}}/C_{\text{Gurobi}} < 1$) for 5/10 of the $(8, 8)$ instances, 6/10 of the $(8, 10)$, 4/10 of the $(8, 12)$, and 14/30 of the $(8, 14)$ instances. All methods have the Make\_Diff method applied to them in postprocessing.}
    \label{fig:Gurobi_performance}
\end{figure*}

\section{Gurobi Solver} \label{app:gurobi}
We formulate an integer linear programming (ILP) of our fleet assignment compatible with Gurobi (and other ILP solvers such as CPLEX and GLPK). This formulation evades the need to enumerate all of the feasible assignments (see Eq.~\eqref{eq:IMP}). 

Let $K$ denotes the set of tours, $C$ the set of vehicles, and $V$ the set of vehicles classes. The incompatibilities between tours are identically represented as edges of the conflict graph $E \subseteq K \times K$, and each tour $k_i$ has a set of allowed vehicle classes $V_{k_i} \subseteq V$. Each vehicle class $v_j$ has a specific operational cost $C^{(v)}_{v_j}$ and each tour $k_i$ has an associated tour cost $C^{(T)}_{k_i}$. We introduce the following binary decision variables:
\begin{equation}
z_{c,v} =
\begin{cases} 
1 & \text{if vehicle } c \text{ is assigned model } v,\\
0 & \text{otherwise,}
\end{cases}
\end{equation}
\begin{equation}
x_{c,k} =
\begin{cases} 
1 & \text{if tour } k \text{ is assigned to vehicle } c,\\
0 & \text{otherwise.}
\end{cases}
\end{equation}
The ILP formulation of the assignment problem then takes the following form.
\begin{equation}
\begin{split}
    \min \quad & \sum_{c \in C} \sum_{v \in V} C^{(v)}_v z_{c,v} + \sum_{c \in C} \sum_{k \in K} C^{(T)}_k x_{c, k} \label{eq:MILP} \\
    \text{s.t.} \quad & \sum_{v \in V} z_{c, v} \leq 1, \quad \forall c \in C   \\
                      & x_{c, k} \leq \sum_{v \in V} z_{c, v} \quad \forall c \in C, \forall k \in K \\
                       z_{c, v} +& x_{c, k_i}  + x_{c, k_j} \leq 1 \quad \forall \begin{cases}
      c \in C \\
     (k_i, k_j) \in E \\
     v \in V \setminus V_{k_i} \intersection V_{k_j}
    \end{cases}   \\
                      & \sum_{c \in C} x_{c, k} \geq 1 \quad \forall k \in K \\
                      & \sum_{c \in C} z_{c, v} \geq N_v^{\min} \quad \forall c \in C \\
                      & \sum_{c \in C } z_{c, v} \leq N_v^{\max} \quad \forall c \in C \\
\end{split}
\end{equation}
The first constraint enforces that each vehicle $c$ is assigned to an unique class $v$. The second constraint enforces the fact that a tour $k$ can be assigned to a vehicle $c$ only if it is available and assigned to a specific class. The third constraint simultaneously enforces that incompatible tours $k_i$ and $k_j$ cannot be assigned the same vehicle $c$ while enforcing that $c$ cannot be assigned to the two tours if its class $v$ is not compatible with the allowed classes of both tours $k_i$ and $k_j$. This distinction between vehicles and vehicle classes is important, because it is possible that two incompatible tours are assigned to separate vehicles of the same class $v$ as long as they are not assigned the same vehicle $c$, hence the need for the set $C$ of vehicles. In the CG formulation, the distinction between vehicle and vehicle classes is implicit in that each column represents one physical vehicle. For example, consider a pool $\Lambda_1 = \{(0, (1)), (0, (2))\}$ consisting of two IS compatible with the same class, versus a pool $\Lambda_2 =\{0, (1, 2)\}$. The first pool assigns separately one vehicle of class $0$ to tour $1$ and another vehicle of the same class to tour $2$, whereas in the second case one vehicle of class $0$ is assigned to both tours. Finally, the fifth constraint enforces that each tour must be assigned to at least one vehicle and the last two constraints enforces the availability bounds on the number of classes. Finally, to guarantee that a feasible solution can always be found, the size of the set of vehicles is bound $\abs{C} = \min ( \sum_{v \in V} N_v^{max}, \abs{K})$, where the first term is a physical upper bound where all the vehicle models are assigned, while the second term corresponds to the special case where each tour is assigned its own vehicle.

We solve the ILP for all our synthetic instances using the Gurobi solver ~\cite{gurobi} (version 12.0.0). Each instance is allocated a maximum runtime of 8 hours and is executed on 5 CPU cores (Intel 6972P @ 2.4 GHz) with a maximum memory of 50GB allowed. We compare, instance by instance, our various CG workflow heuristics with Make\_Diff post-processing against Gurobi in fig~\ref{fig:Gurobi_performance}. Across these runs, Gurobi produced superior solutions in 5 out of 10 of the $(8, 8)$ instances, 6 out of 10 of the $(8, 10)$ instances, 4 out of 10 of the $(8, 12)$ instances, and 14 out of 30 of the $(8, 14)$ instances. Gurobi, under the time and memory constraints given, is thus unable to return a reliably better solution than our CG scheme. This is likely due to it getting stuck in local minima during its branch-and-bound exploration. This explains why we use ILP+DIV in our benchmark in the main text.

\bibliography{ref-jabref}

\end{document}